\documentclass[useAMS,usenatbib]{mn2e}
\usepackage{xcolor}
\usepackage{graphicx}
\usepackage[T1]{fontenc}
\usepackage{aecompl}

\title[3D map using the Lyman-alpha forest]
  {Nonparametric 3D map of the IGM using the Lyman-alpha forest}
 \author[Cisewski et al.]
   {Jessi Cisewski,$^1$ 
Rupert A. C. Croft,$^2$ 
Peter E. Freeman,$^1$ 
Christopher R. Genovese,$^1$ \newauthor
Nishikanta Khandai,$^3$ 
Melih Ozbek,$^2$ 
Larry Wasserman$^1$ \\
   $^1$ Department of Statistics, Carnegie Mellon University, Pittsburgh, PA, USA \\
   $^2$ Department of Physics, Carnegie Mellon University, Pittsburgh, PA, USA \\
   $^3$ Department of Physics, Brookhaven National Laboratory, Upton, NY, USA}
\date{Released 2002 Xxxxx XX}

\pagerange{\pageref{firstpage}--\pageref{lastpage}} \pubyear{2002}

\def\LaTeX{L\kern-.36em\raise.3ex\hbox{a}\kern-.15em
    T\kern-.1667em\lower.7ex\hbox{E}\kern-.125emX}

\begin{document}

\label{firstpage}

\maketitle

\begin{abstract}

Visualizing the high-redshift Universe is difficult due to the dearth of available data; however, the Lyman-alpha forest provides a means to map the intergalactic medium at redshifts not accessible to large galaxy surveys.  Large-scale structure surveys, such as the Baryon Oscillation Spectroscopic Survey (BOSS), have collected quasar (QSO) spectra that enable the reconstruction of HI density fluctuations.  The data fall on a collection of lines defined by the lines-of-sight (LOS) of the QSO, and a major issue with producing a 3D reconstruction is determining how to model the regions between the LOS.  We present a method that produces a 3D map of this relatively uncharted portion of the Universe by employing local polynomial smoothing, a nonparametric methodology.  The performance of the method is analyzed on simulated data that mimics the varying number of LOS expected in real data, and then is applied to a sample region selected from BOSS.  Evaluation of the reconstruction is assessed by considering various features of the predicted 3D maps including visual comparison of slices, PDFs, counts of local minima and maxima, and standardized correlation functions.   This 3D reconstruction allows for an initial investigation of the topology of this portion of the Universe using persistent homology.  
\end{abstract}


\section{Introduction} 

A high fidelity 3D map of the high redshift Universe would be a useful cosmological tool.  It could, for example, aid in ruling out cosmological theories or refining Baryon Acoustic Oscillation (BAO) measurements.  A prominent obstacle in the ability to reconstruct a map is that distant objects must be very luminous for us to observe them.  Fortunately, bright quasars (QSOs) provide a means for probing the HI density fluctuations of the Intergalactic Medium (IGM) in unexplored regions via the Lyman-alpha forest.  HI density fluctuations in the IGM can be inferred by analyzing the Lyman-alpha absorption features in QSO spectra (\citealt{Lynds1971}). The Lyman-alpha absorption is generated by a diffuse and continuous medium 
which traces the dark matter density structures on scales
above a pressure smoothing scale of $\sim$ 0.1 h$^{-1}$Mpc 
(e.g., see the review by \citealt{Rauch2006}). It can therefore be
used to paint a faithful picture of the Universe on larger scales. 
Potential uses of the Lyman-alpha forest data include measuring clustering on large-scales (\citealt{SlosarEtAl2011}), placing constraints on inflation (\citealt{palanque2013}), and measuring BAO (\citealt{busca2013, slosar2013}).  Additionally, to get a better idea of the matter distribution of the IGM it is necessary to understand its properties between the QSOs' lines of sight (LOS).

Previous attempts have been made to combine the 1D data in individual Lyman-alpha forest sight lines to develop a 3D picture of the HI density. In particular a Wiener interpolation was proposed and tested with data from cosmological simulations by \cite{PichonVergelyEtAl2001} and \cite{CaucciColombiEtal2008}  
(see also recent work by \citealt{lee2013}).  We propose a different, nonparametric statistical methodology for producing a 3D map of high-redshift HI density fluctuations using local polynomial smoothing (\citealt{Wasserman2006}).  The method is tested here on simulated data that mimics the varying number of LOS expected in the real data, and applied to a sample region selected from the Baryon Oscillation Spectroscopic Survey (BOSS) (\citealt{dawson2013}).  We evaluate the predicted 3D maps by considering various summaries of the maps including visual comparison of slices, PDFs, counts of local minima and maxima, and standardized correlation functions.   Decreasing the number of LOS used in the estimation procedure will naturally result in depreciated performance so rather than solely comparing the 3D reconstructed maps with varying LOS to the map derived from the full simulated dataset, additional comparisons are used to provide a clearer picture of the information lost when fewer LOS are available.  This 3D reconstruction provides a means for an initial investigation of the topology of this portion of the Universe using persistent homology (\citealt{Sousbie2011,SousbieEtAl2011, WeygaertEtAl2013}) of the reconstructed maps.  Finally, we apply the proposed methodology to a sample portion of the data from the Lyman-alpha forest in BOSS SDSS Data Release 9 (\citealt{LeeEtAl2013}).  

In {\S}2, we review the properties of the Lyman-alpha forest, then in {\S}3 we provide an introduction to local polynomial smoothing.  We analyze the performance of the proposed methodology on simulated data {\S}4 followed by an application to a subset of the BOSS data in {\S}5.  Finally, in {\S}6 we provide some closing remarks.

\section{Lyman-alpha forest}\label{sec:data}
The Lyman-alpha forest provides a means of studying the large-scale structure in the redshift range  $ 2 \leq z \leq 3.5 $ that is complementary to galaxy survey studies at low redshift. This quoted redshift range is limited on the low-redshift end by the atmosphere's absorption of ultraviolet light and for our current purposes on the high redshift end by the rarity of bright quasars.

 Due to the expansion of the Universe, the wavelength of the photons in QSO spectra increases by a factor of $(1+z)$, where $z$ is the relative redshift between two points in space. Along the LOS from the Earth to a QSO, photons with a wavelength of 1216 {\AA } can be absorbed by HI, which can undergo a transition from the ground state to the first excited state. This 
absorption is an imprint of the location and the density of the HI. Since hydrogen is assumed to be in photoionization equilibrium in the moderate density IGM
that gives rise to the forest, one can then infer the total baryon density (\citealt{rauch1997}). Also, as baryons follow the total matter potential, it is a tracer of matter, including dark matter and one can study large-scale structure (see e.g., \citealt{mcdonald2003}).

Until recently, data from Lyman-alpha surveys larger than a single close pair of QSOs has usually been treated as a collection of 1D problems for individual QSO sightlines.  Now however, the high areal density of QSOs in current surveys is making it possible to correlate information three dimensionally
(see \citealt{SlosarEtAl2011}). BOSS, which is one of the four surveys of the SDSS III (\citealt{eisenstein2011}), features a QSO density of at least 15 $\deg^{-2}$ ($\sim150,000$ QSOs over 10,000 $deg^{2}$). Each QSO provides Lyman-alpha forest information along a skewer of comoving length $\sim$400 h$^{-1}$Mpc.  The mean comoving separation between QSO spectra in BOSS is $\sim$20 h$^{-1}$Mpc. We test our methods using a small sample of  BOSS Data Release 9 (DR 9) in this paper to show that they can be applied to real data as well as simulations. In the future, one can expect an even better recovery of the field because even more QSOs will be available for analysis (e.g., 45 $\deg^{-2}$ proposed for MS-DESI, \citealt{schlegel}).

\section{Methodology}
The main goal of the proposed methodology is to produce a 3D map of the IGM using the Lyman-alpha forest.  However, predicting the HI density fluctuations in the IGM between the LOS of the QSOs is a delicate problem due to the sampling scheme entailed by the LOS and the location of the QSOs.  Any statistical methodology will naturally require some assumptions on the intervening regions; given the relatively uncharted state of the IGM, the desire is to keep the assumptions minimal.  The proposed method for producing a 3D map of the IGM is local polynomial smoothing, which has a number of benefits noted below.  An introduction to local polynomial smoothing, along with details regarding its implementation, can be found in \cite{ClevelandGrosseShyu1992}, or see \cite{Wasserman2006} for an overview of nonparametric statistical methods.  Assume that we want to estimate an unknown function $f$
    but only observe noisy samples of the function $Y_1, Y_2, \ldots, Y_n$ at discrete points
    $X_1, \ldots, X_n$. The data can be described by the model
    \begin{equation}
    Y_i = f(X_i) + \epsilon_i, \quad i = 1, \ldots, n,
    \end{equation}
    where the $\epsilon_i$ are independent random errors with expectation 0.
    For example, with the Lyman-alpha forest data, the $Y_i$ are the measured
    HI density fluctuations at points in space $X_i$ (represented by RA, DEC, and redshift),
    and $f$ represents the unknown map of HI density fluctuations across a given region 
    of space.

The general idea of local polynomial regression is to estimate the function $f$ locally using a $d$-degree polynomial.  Estimation is local in the sense that only observations within a neighborhood are used for estimation, which requires specification of the neighborhood size.  The neighborhood size is controlled by a smoothing parameter $\alpha$, which lies between 0 and 1, indicates the portion of the full data set used in estimating the function at a given point.  (Note that the smoothing parameter can also be defined by the distance from the observations.)  The number of observations used in the local fit is $n_{\alpha} = \lfloor\alpha n\rfloor$, the largest integer less than or equal to $\alpha n$.  A larger $\alpha$ gives a smoother fitted function.  A benefit of this type of bandwidth selection is that, through $\alpha$,  it adapts for unevenly spaced explanatory variables:  the neighborhood becomes larger in regions with fewer observations and smaller in regions with more observations.  This is especially important for fitting the BOSS QSO data since the location of the QSOs are not evenly spaced.

The local polynomial estimator $\hat f(x)$\footnote{Estimates are denoted with a { $\hat{ }$ } so $\hat{g}$ is an estimate of the function $g$.} at some point $x$ is achieved by finding $\hat{a} = (\hat{a}_0, \hat{a_1}, \ldots, \hat{a}_d)^T)$ that minimizes
$$ 
\sum_{i = 1}^n \left(Y_i - P_x(x_i; a)\right)^2 K\left(\frac{x_i - x}{h_{\alpha}}\right)
$$
where $P_x(x_i; a) = a_0 + a_1(x_i - x) + \frac{a_2}{2!}(x_i - x)^2 + \cdots + \frac{a_p}{p!}(x_i - x)^d$, $x_i$ for $i = 1, \ldots, n$ are the explanatory variables, and $h_{\alpha}$ is the bandwidth where the subscript indicates its dependence on $\alpha$.  $K$ is a kernel function used so that observations that are further away are given lower weight than the observations near the point of interest.  The choice of kernel function does not have a strong influence on the fit of the model (\citealt{FanEtAl1997}).  The estimate at $x$ is then $\hat{f}(x) = P_x(x,\hat{a}) = \hat{a}_0(x)$.  If the degree of the polynomial is $d = 0$, the result is the Nadaraya-Watson kernel estimator (\citealt{HastieLoader1993, Wasserman2006}).

The most common nonparametric regression
method is the Nadaraya-Watson kernel estimator noted above, which takes the form
$\hat f(x) = \sum_i Y_i K((x_i -x)/h) / \sum_i K((x_i -x)/h)$.
However, the kernel estimator has problems in our setting.
Let ${\cal X}$ be the set of $x$ values over which we want to estimate $f(x)$.
In our case, because the observations fall on lines,
the observable $x$'s fall on a subset
${\cal X}_0\subset {\cal X}$.
It turns out that
the kernel estimator has considerable bias
at boundaries of ${\cal X}_0$.
In fact, in our case, all of ${\cal X}_0$ is a non-interior set of ${\cal X}$.
This means that kernel smoothing would suffer large bias everywhere.
Local polynomial smoothing with $d\ge 1$
does not suffer from boundary bias (\citealt{FanGijbels1992, HastieLoader1993, RuppertWand1994}).
Hence, local polynomial smoothing is preferred in this setting.

Local polynomial smoothing is a useful methodology for modeling the IGM using Lyman-alpha forest data for several reasons.  As previously noted, the natural adaption of the smoothing parameter to unevenly spaced explanatory variables is important since  QSOs are not observed on an evenly spaced grid.  Another benefit is that no global assumption about the data need to be made since the fit is done locally.  Furthermore, by considering polynomials of degree $d \geq 1$, the design and boundary bias present in kernel regression is removed (\citealt{FanGijbels1992, HastieLoader1993, RuppertWand1994}).  Removing the design bias is particularly important given the design of the Lyman-alpha forest data as approximately parallel lines in a three-dimensional region (i.e. the data falling on almost parallel lines results in a design bias; the design is the placement of the spatial points where HI density fluctuation is measured).  Another useful feature of local polynomial smoothing is that standard errors of the estimates are available, which provides a means of representing the uncertainty in the predicted map. (For an illustration, see Figure~\ref{fig:slice1000}, which  displays a predicted layer of the simulated data plus or minus the standard errors.)

\section{Analysis of Simulated Data}
In order to evaluate the expected performance of local polynomial smoothing
on \
Lyman-alpha forest data from BOSS and other observational surveys, we made
use
of a large hydrodynamic cosmological simulation of the $\Lambda$CDM model.
We used the smoothed particle hydrodynamics code {\small{P-GADGET}}
(see \citealt{springel,dimatteo2005})
to evolve a distribution of $2\times4096^{2}=137$ billion particles in a
cubical periodic volume of side-length 400 h$^{-1}$Mpc (\citealt{Khandai2014}).
The simulation cosmological parameters were $h=0.702$, $\Omega_{\rm m}=0.275$,
$\Omega_{\rm \Lambda}=0.725$,  $\Omega_{\rm b}=0.046$, spectral index $n_s$ = 0.968, and amplitude of mass fluctuations,
$\sigma_{8}=0.82$.
The mass per particle was $1.19\times10^{7} h^{-1}M_{\odot}$
 (gas) and $5.92\times10^{7} h^{-1}M_{\odot}$ (dark matter). A
gravitational force resolution of $3.25$ h$^{-1}$kpc comoving was used.
The power spectrum of the simulation initial conditions was taken from
CAMB (\citealt{camb}).
The simulation was run with
an ultraviolet background radiation field consistent with \cite{hm96}.
Cooling
and star formation were included. However the latter used a lower density
threshold than usual (for example in \citealt{springelh03})
so that gas particles are rapidly converted
to collisionless gas particles. This was done to
speed up execution of the simulation
(see \citealt{Khandai2014} for more details).
As a result the stellar properties of galaxies in the simulation are not
predicted reliably but this has no significant effect on the diffuse IGM
that gives rise to the Lyman-alpha forest.
Black hole formation and feedback from stars were also switched off in the
simulation.

The simulation output at redshift $z=2$ was used to generate a grid
of Lyman-alpha spectra (see \citealt{hernquist})
 with $176^2=30,976$ evenly spaced sightlines 
(resulting in 2.27 h$^{-1}$Mpc spacing).  Each sightline was generated with high resolution,
10,560 pixels, in order to resolve the thermal broadening when computing
the optical depth. They were then down-sampled (by averaging the
transmitted  flux over 60 pixels) to 176 pixels.  The full set of simulation data therefore consists
of $176^{3}$ data values, which will be referred to as the high-resolution 
dataset.

The cosmological simulation measurement at each grid location, $(X, Y, Z)$, is the delta flux, $\delta_i = \frac{e^{-\tau_i}}{<e^{-\tau}>} - 1$, where $\tau_i$ is the optical depth at location $(X_i,Y_i,Z_i)$ for $i = 1, \ldots, n$ where $n = 176^3$ (the number of grid locations), and $<e^{-\tau}> = n^{-1} \sum_{i = 1}^n e^{-\tau_i}$.  \emph{Higher} values of $\delta$ therefore correspond to \emph{lower} density regions and \emph{lower} values of $\delta$ correspond to \emph{higher} density regions.

\begin{figure}
  \centering
      \includegraphics[width=0.5\textwidth]{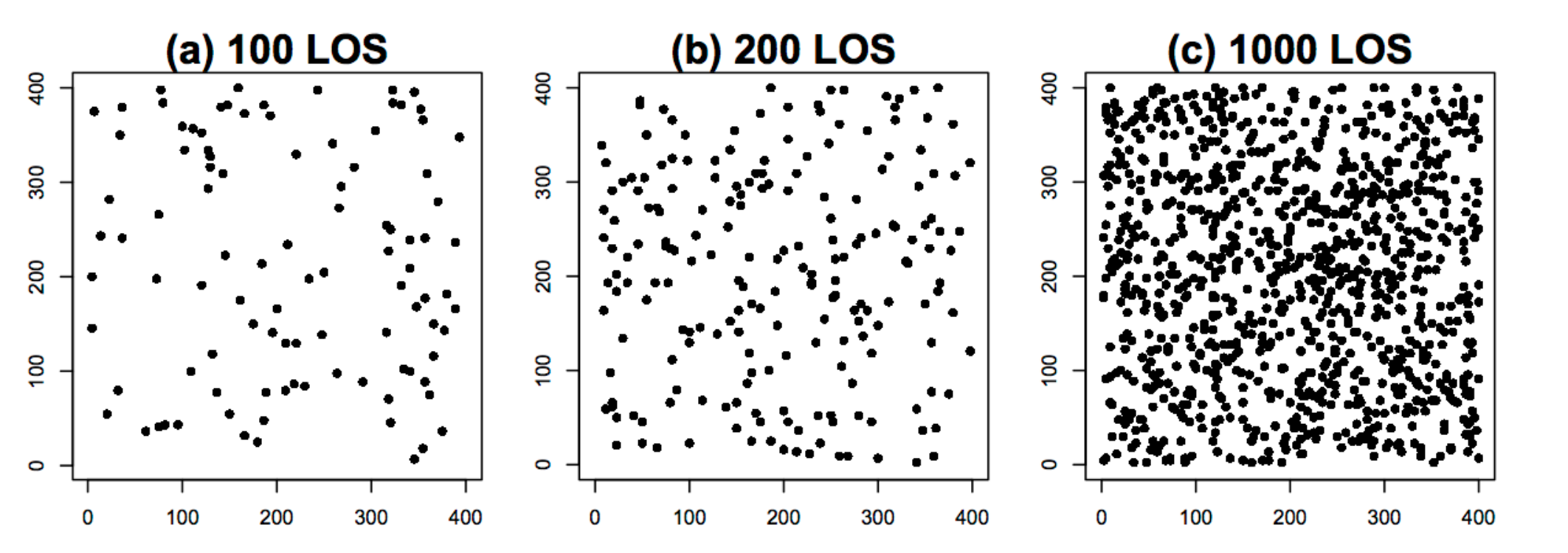} 
  \caption{Simulation designs for the varying number of LOS:  (a) 100 LOS, (b) 200 LOS, (c) 1000 LOS.  The horizontal and vertical axes correspond to RA and DEC, and the redshift direction would be perpendicular to the page.}
\label{fig:simulation_design}
\end{figure}

The effective angular density of number of QSO LOS in the simulation cube is significantly higher than that in real data.  For example, in the BOSS survey Lyman-alpha forest dataset (\citealt{LeeEtAl2013,OzbekEtAl2014}) the average number of sightlines passing through an area of (400 h$^{-1}$Mpc)$^2$ is between $\sim 100$ (at redshift $z=3$) and $\sim 400$ (at redshift $z=2$). In order to mimic the design of the real Lyman-alpha forest data which has fewer LOS and quasars locations not aligned with a grid, we randomly selected a sample of 100, 200 and 1000 LOS from the high-resolution dataset.  Note that the 100, 200, and 1000 LOS datasets are \emph{subsets} of the high-resolution data; different cosmological simulations were not used for each new sample.  Figure~\ref{fig:simulation_design} displays the $X$ and $Y$ locations of these samples.  

For each sample, the smoothing parameter, $\alpha$, was selected using generalized cross validation (GCV) (\citealt{Wasserman2006}) with the tri-cubic weight function (\citealt{FanEtAl1997, Wasserman2006}).  After each sample was modeled using local polynomial smoothing, the resulting model was used to predict the values on the high-resolution grid with $176^3$ locations.  In order to assess the performance of the proposed methodology, various summaries of the predicted grids were compared:  (i) visually comparing individual slices at a fixed redshift, (ii) plotting PDFs (smoothed histograms of the predicted values), (iii) counting the number of local minima and local maxima, and (iv) determining a standardized correlation function, which are all explained in more detail below.  However, it is expected that decreasing the number of LOS used in the modeling procedure will result in inferior performance compared to datasets with a higher number of LOS; that is, the predicted map using the 100 LOS dataset is not expected to perform as well as the full, high-resolution dataset nor is it expected to capture all the features of the full dataset.  

An improved method for assessing the performance of the local polynomial smoothing in this setting is to compare the predicted regions using the 1000, 200, and 100 LOS with the predicted regions using the same smoothing parameter values, $\alpha$, applied to the full, high-resolution data.  The latter predictions are designated 1000hr, 200hr, and 100hr where the ``hr'' stands for ``high-resolution'' indicating that the particular LOS $\alpha$ was used on the full high-resolution dataset.  Though this mode of comparison reduces the degree-of-smoothing bias over comparing the LOS predictions directly to the high-resolution predictions, issues remain and are discussed below.

The measures of assessment noted above are important to consider because they can reveal deficiencies in an employed methodology; however, none of the measures describe the topological features of the data such as its hole structure.  The topological data analysis tool called persistent homology provides a useful framework for characterizing these important features.  We use persistent homology to heuristically compare the topology of the predicted regions as the available number of LOS changes.  In particular, a comparison of the persistence diagrams (introduced below) reveal which topological features are lost when reducing the number of LOS.

Computations were performed using the statistical software R (\citealt{R2012}), and, in particular, the function \emph{loess} was used for the local polynomial regression.

\subsection{Smoothing parameter selection}
\begin{figure}
  \centering
     \includegraphics[width=2.5in]{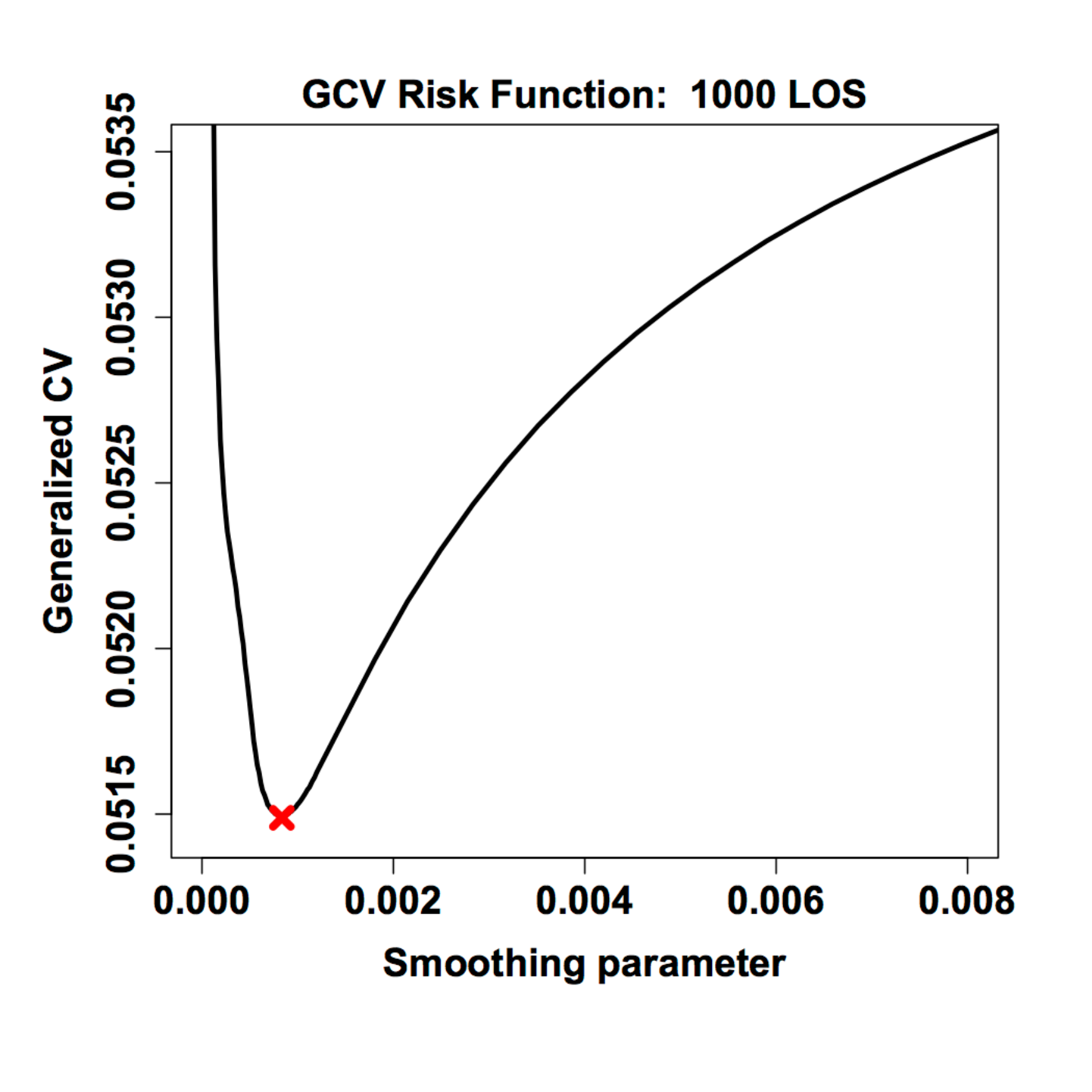} 
  \caption{The risk function for selecting the smoothing parameter of the 1000 LOS data.  The smoothing parameter determines the size of the neighborhood in the estimation procedure.  Generalized cross validation (GCV) was used, and the goal is to minimize the GCV error.  The minimum occurs around $0.00083$ marked by the red `x'.} 
\label{fig:gcv}
\end{figure}

The first step in employing local polynomial smoothing for the simulated data is to select the smoothing parameter, $\alpha$, for each dataset.  A larger $\alpha$ means more data will be used in the local fit, which results in a smoother surface (and a smaller $\alpha$ means fewer observations will be used in the local fit, which results in a rougher surface).  The ``correct'' amount of smoothing for a given dataset depends on several factors including the goal of the analysis.  We selected the smoothing parameter for each dataset using generalized cross validation (GCV), and the values are displayed in Table~\ref{tab:smoothing}. GCV approximates the  predictive risk of the local polynomial fit, and the goal is to select the $\alpha$ that minimizes the GCV criterion.  The general idea of cross validation is to leave out one or more observations in the estimation, then compare how the estimated function performs on the observations that were left out.  As an example, Figure~\ref{fig:gcv} displays the estimated risk function for the 1000 LOS dataset. For the 200 and 100 LOS datasets, the $\alpha$ that minimized GCV produced a neighborhood that was not sufficient for estimation so the smallest $\alpha$ that allowed for estimation was used.  As the sample size, $n$, increases, the smoothing parameter value tends to 0 at a rate of $n^{1/7}$ (\citealt{FanGijbels1996}). This fact, along with the mean smoothing parameter value for 10 randomly selected datasets with 1000 LOS, was used to determine the $\alpha$ for the full high-resolution dataset.

\begin{center}
\begin{table}
\begin{tabular}{ccc}
{\bf Sample} & {\bf Smoothing parameter} & {\bf $N$} \\
\hline
100    &  0.02174 & 383 \\
200   &  0.00836 & 294 \\
1000   &  0.00083 & 146 \\
High-res   &  0.000514 & 2,802 \\
\hline
\end{tabular} \caption{Selected smoothing parameter values using GCV along with the number of observations, $N$, used to estimate the parameters for the local polynomial fit. } \label{tab:smoothing}
\end{table}
\end{center}

\subsection{Estimated model comparison} \label{sec:comparisons}
Local polynomial smoothing provides a channel for producing a 3D map of the IGM using Lyman-alpha forest data.  In this section, we explore different ways of assessing its performance in the proposed setting; this is accomplished by summarizing the predicted maps in several ways. Recall that all the datasets are derived from the same cosmological simulation; however, the varying number of LOS were randomly selected.  This allows for direct comparison between the full high-resolution dataset and the 1000, 200, and 100 LOS datasets revealing how the predicted maps change as the number of LOS decreases.

The predictions were compared by summarizing the 3D predicted maps for each dataset in the following ways:  (i)  visually by slices at a fixed $Z$, (ii) estimated PDFs of the predicted values, (iii) calculating the count of local minima and maxima, (iv)  analyzing the behavior of standardized correlation functions, and (v)  looking at the multiscale topological signatures via persistent homology.  Simulation grid locations are denoted by $(X,Y,Z)$ where $X$ and $Y$ are analogous to RA and DEC, and $Z$ is analogous to redshift.

As previously noted, it is not expected that the datasets with fewer LOS will have predictions that capture the same degree of detail as the datasets with more LOS; fewer LOS should result in fewer resolved features.  It is imperative to review the performance with decreasing LOS because the real data will not have the high density of the LOS that is available with the high-resolution dataset (\citealt{ParisPetitjeanEtAl2012}).  Furthermore, while comparing the 1000, 200, and 100 LOS predictions to the high-resolution predictions is helpful in gauging the information lost with fewer LOS, a better - though not perfect -  comparison for the LOS predictions is to the 1000hr, 200hr, and 100hr LOS predictions. 

\begin{figure}
 \includegraphics[height=2.25in]{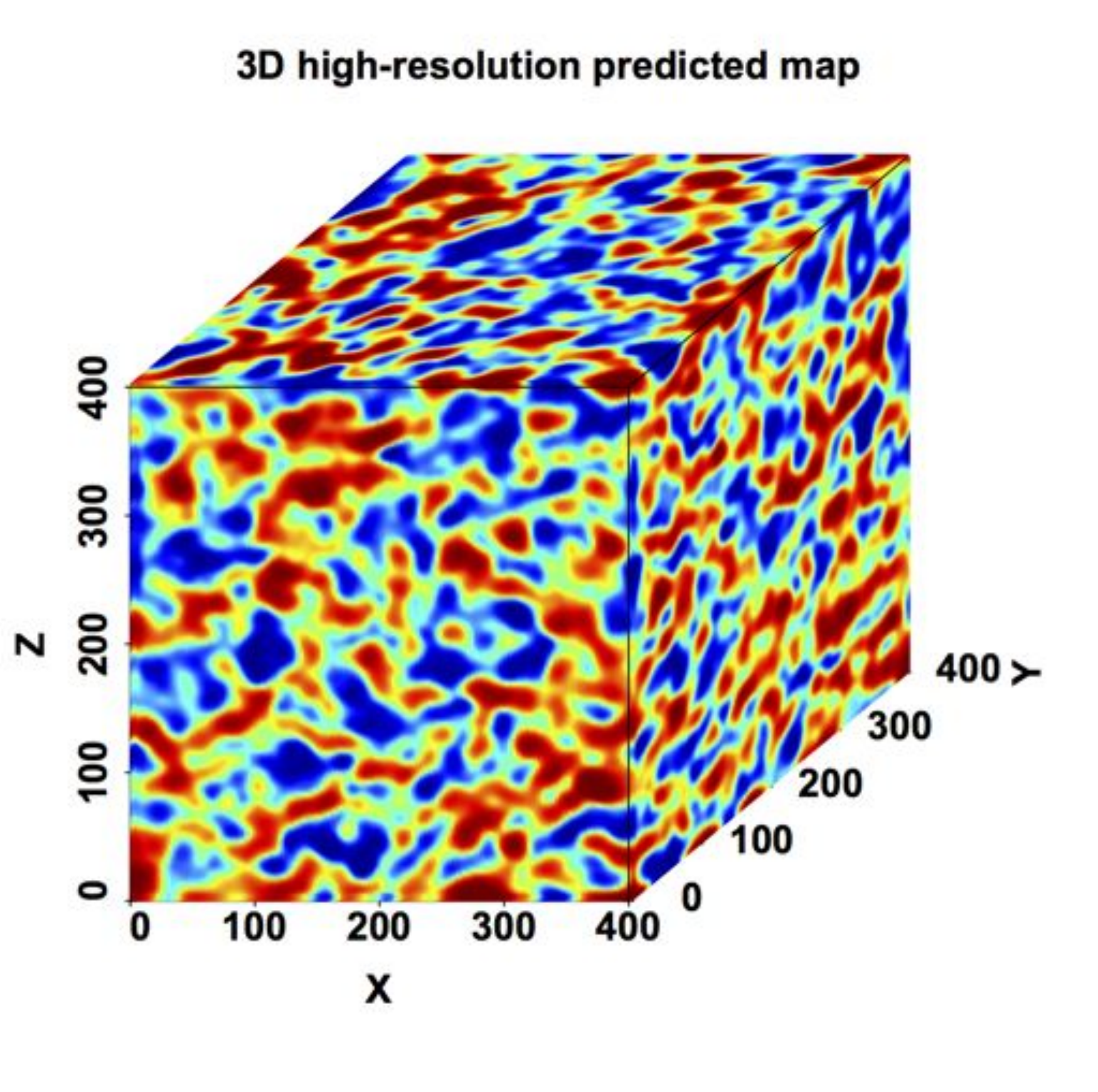}    \includegraphics[height=2in]{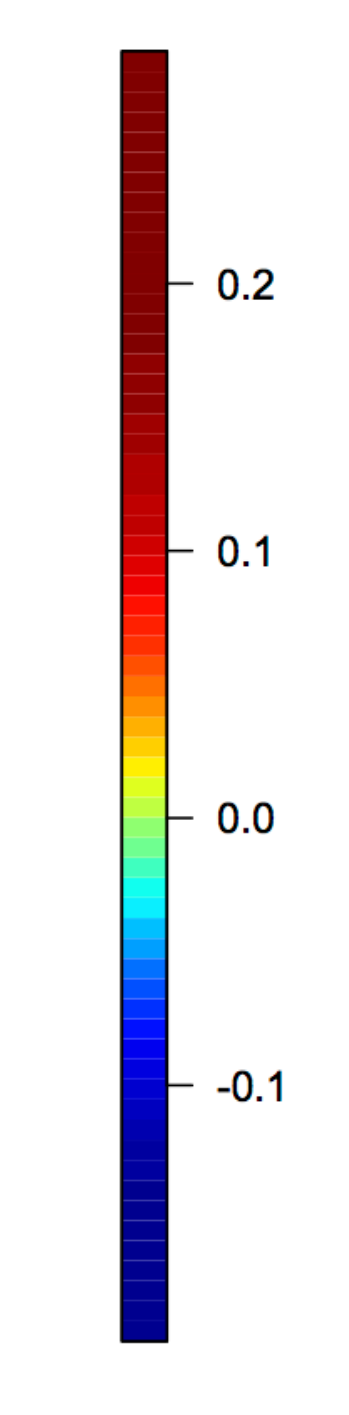} 
  \caption{A 3D image of the simulation cube using the full, high resolution dataset.  Note that the red corresponds to higher density regions (lower flux) while the blue corresponds to lower density regions (higher flux).  The specific values are the negative of the delta flux ($-\delta$). }\label{fig:cube_hr}
\end{figure}

The smoothed, predicted maps for the various datasets will be referred to as $S_{hr}$, $S_{1000}$, $S_{1000hr}$, $S_{200}$, $S_{200hr}$, $S_{100}$, and $S_{100hr}$.  The 3D map of $S_{hr}$ is displayed in Figure~\ref{fig:cube_hr}.

\begin{figure*}
  \centering
   \includegraphics[height=2.5in]{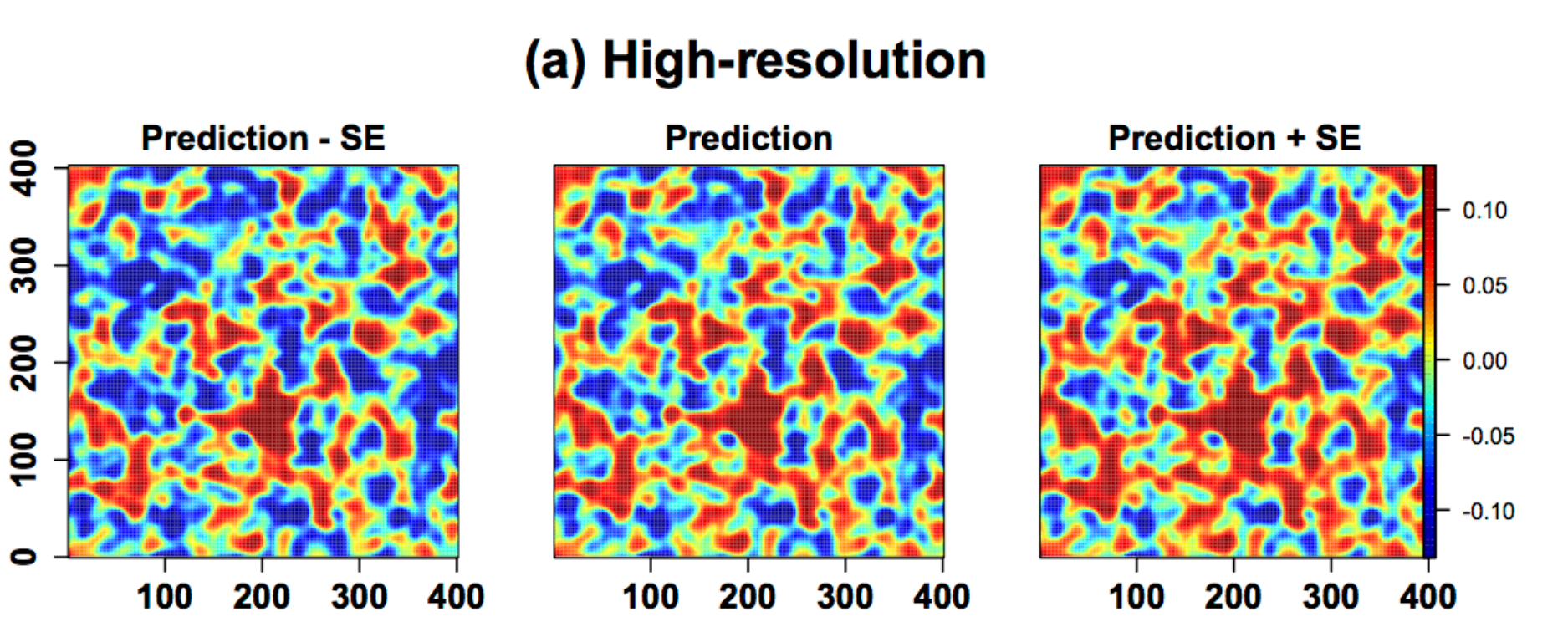} \\
  \includegraphics[height=2.5in]{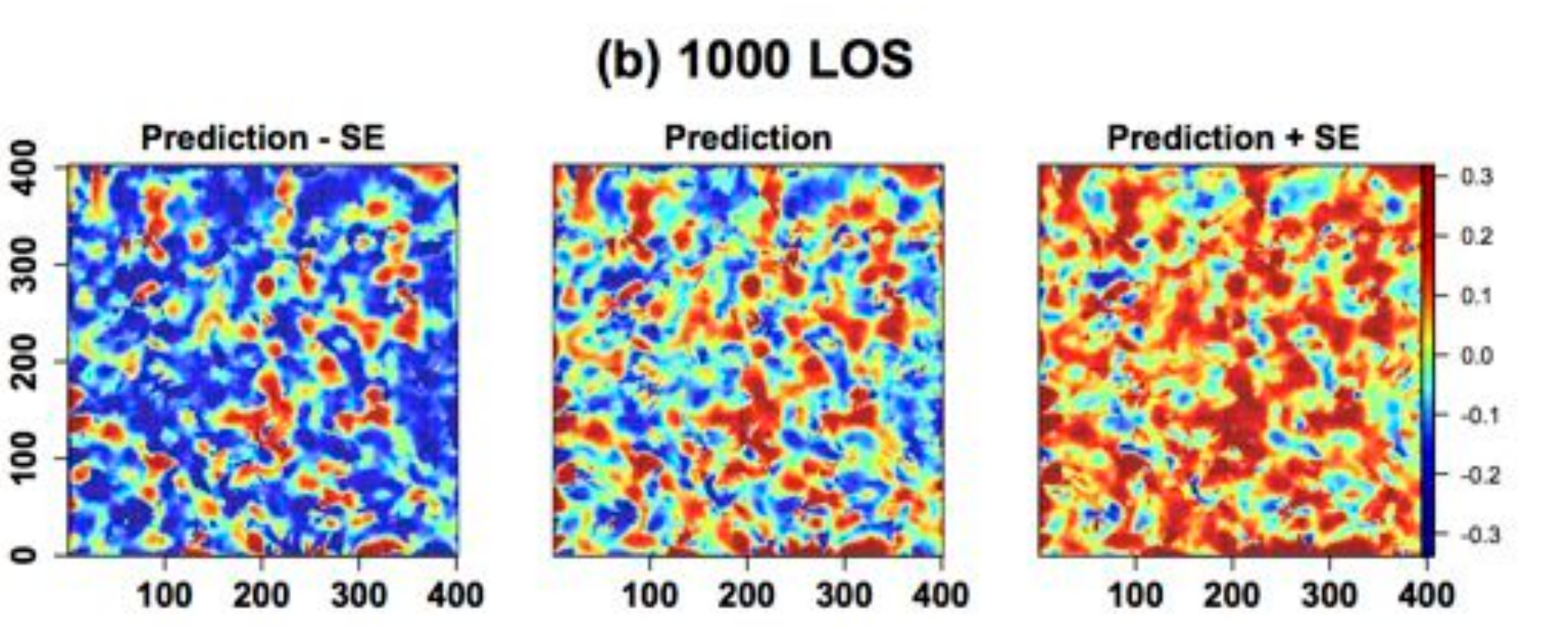} \\
  \includegraphics[height=2.5in]{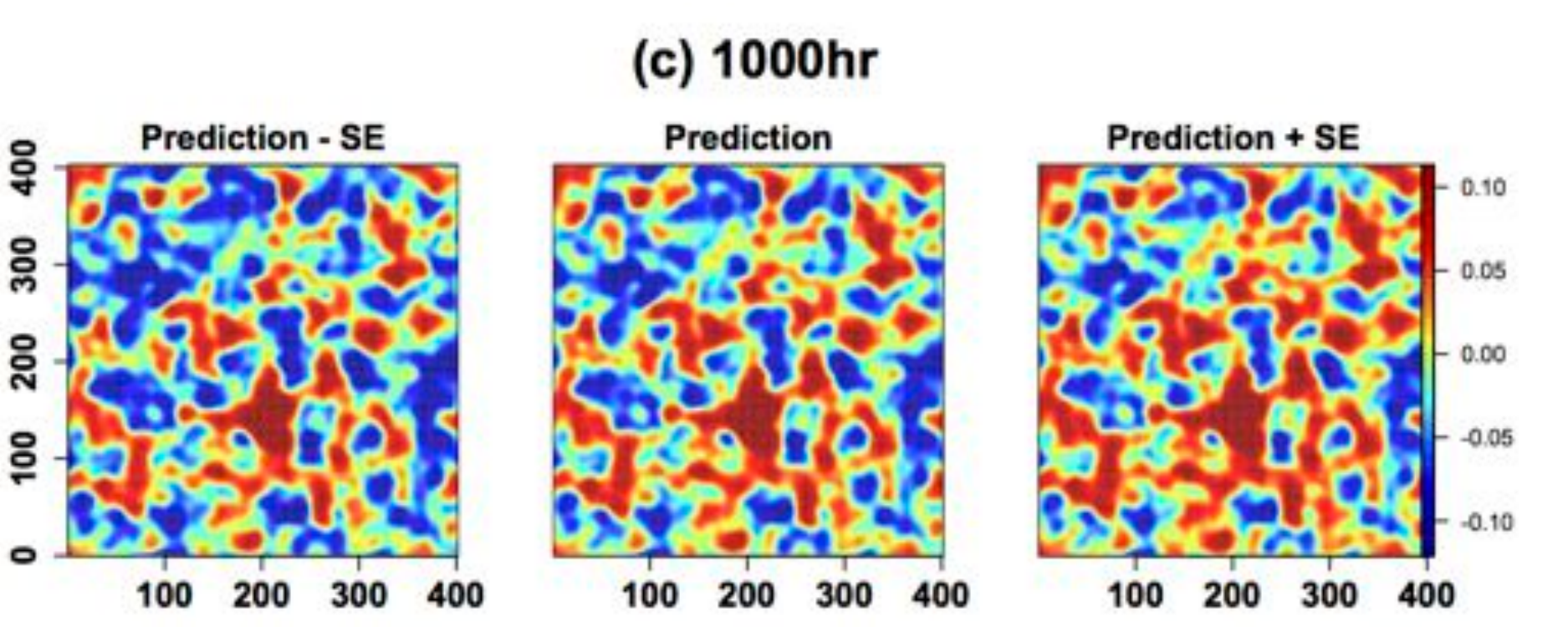} 
  \caption{Slices of the predicted values of (a) the high-resolution data, (b) the 1000 LOS data, and (c) the full high-resolution data using the bandwidth selected for the 1000 LOS dataset (1000hr).  The figures in the first column are the predicted values minus standard error (SE); the middle column figures are the predicted values, and the third column figures are the predicted values plus SE.  Note that the red corresponds to higher density regions (lower flux) while the blue corresponds to lower density regions (higher flux).  The specific values are the negative of the delta flux ($-\delta$). }\label{fig:slice1000}
\end{figure*}

\begin{figure*}
  \centering
  \includegraphics[height=2.5in]{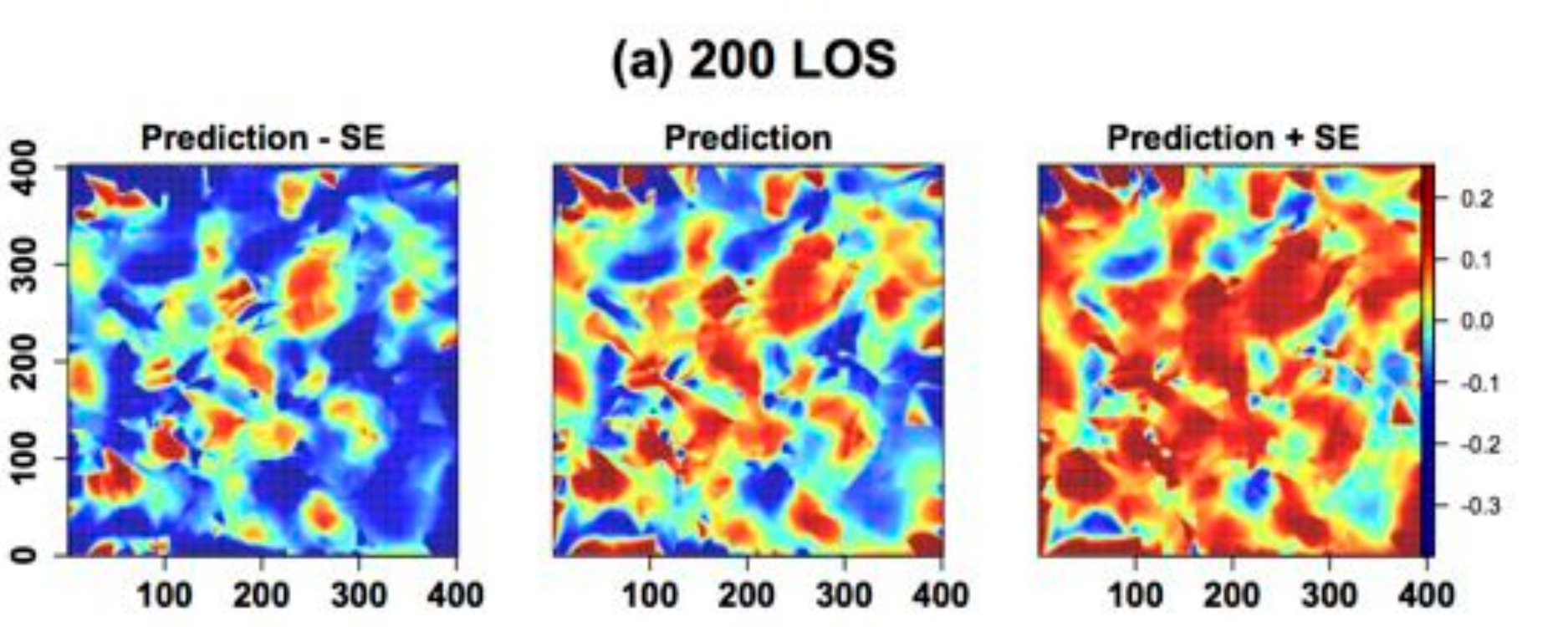} \\
  \includegraphics[height=2.5in]{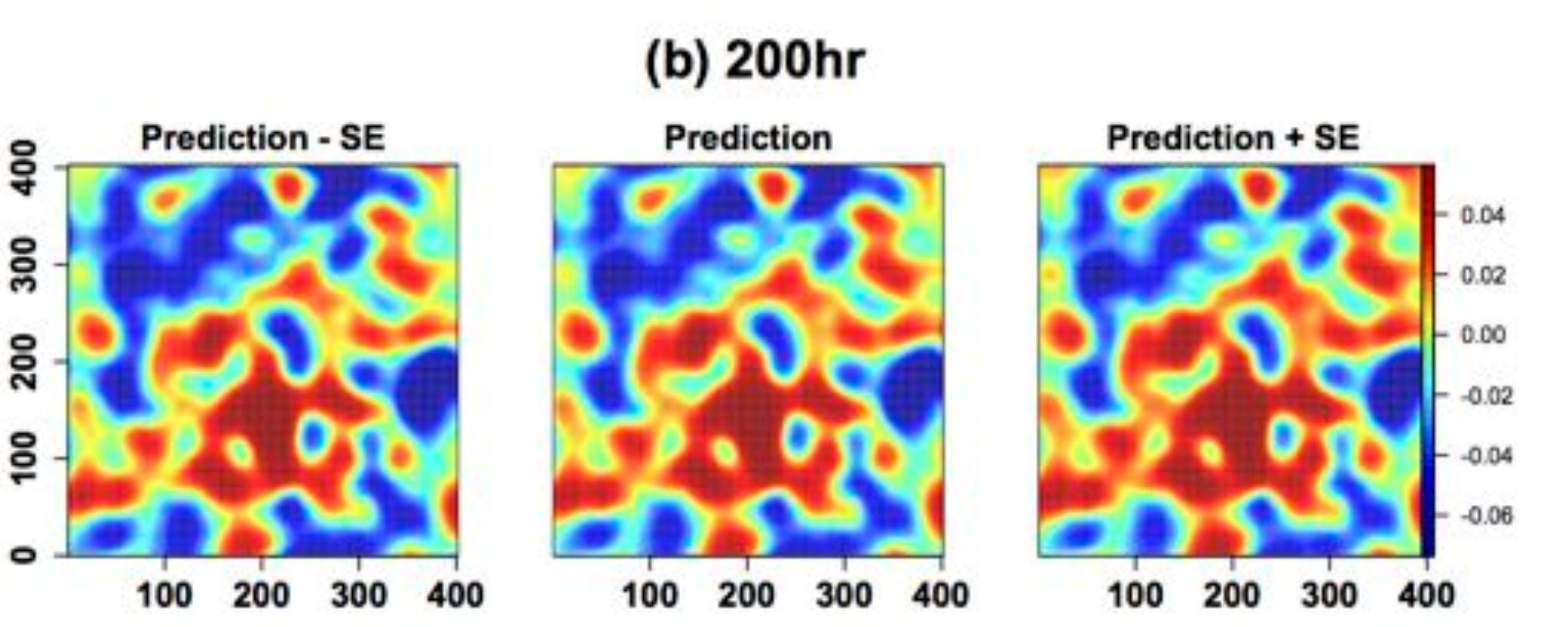} 
  \caption{Slices of the predicted values of (a) the 200 LOS data, and (b) the full high-resolution data using the bandwidth selected for the 200 LOS dataset (200hr).  The figures in the first column are the predicted values minus standard error (SE); the middle column figures are the predicted values, and the third column figures are the predicted values plus SE.  Note that the red corresponds to higher density regions (lower flux) while the blue corresponds to lower density regions (higher flux).  The specific values are the negative of the delta flux ($-\delta$). }\label{fig:slice200}
\end{figure*}

\begin{figure*}
  \centering
  \includegraphics[height=2.5in]{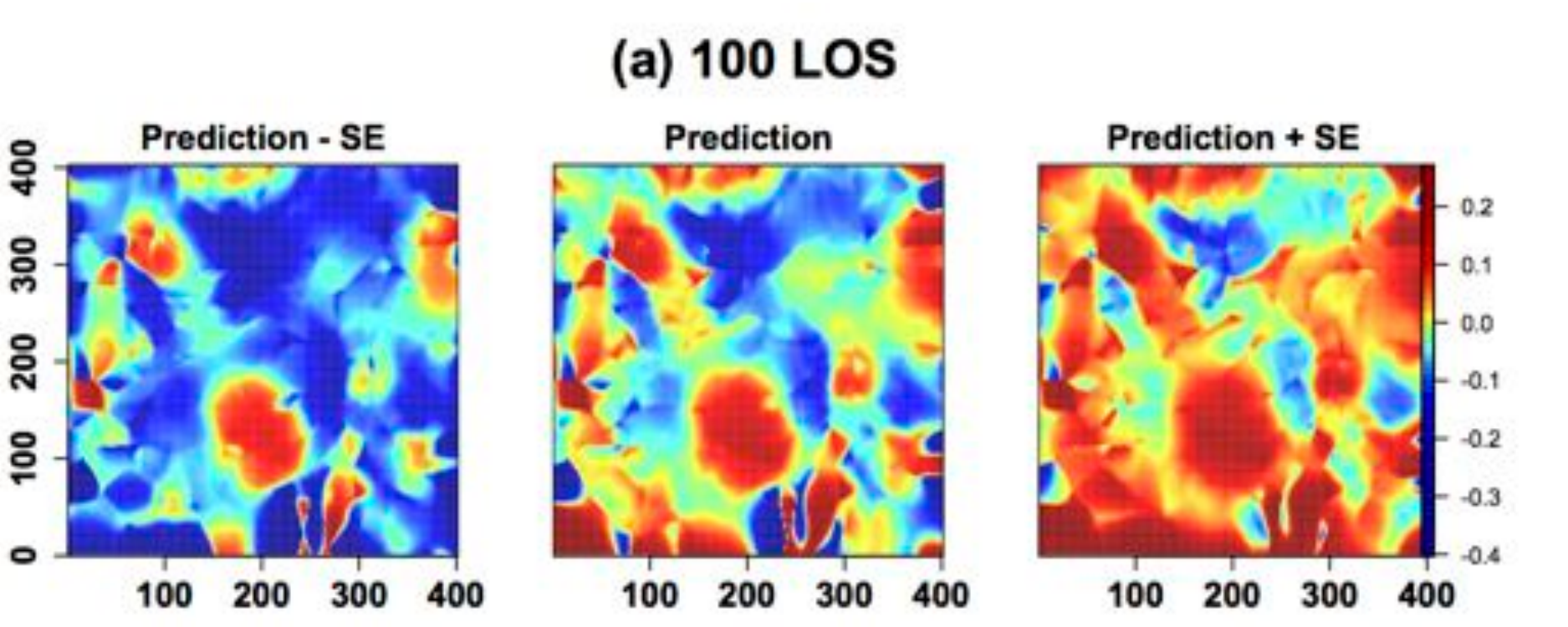} \\
  \includegraphics[height=2.5in]{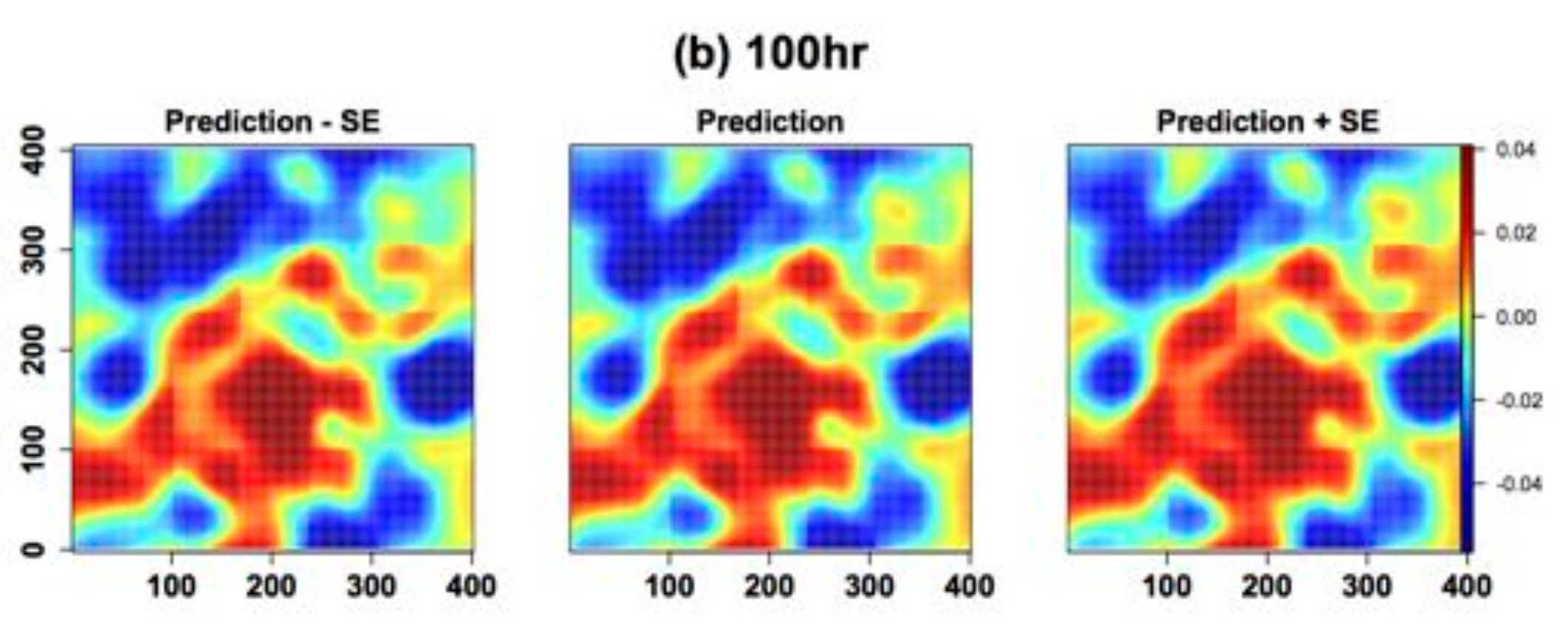}  
  \caption{Slices of the predicted values of (a) the 100 LOS data, and (b) the full high-resolution data using the bandwidth selected for the 100 LOS dataset (100hr).  The figures in the first column are the predicted values minus standard error (SE); the middle column figures are the predicted values, and the third column figures are the predicted values plus SE.  Note that the red corresponds to higher density regions (lower flux) while the blue corresponds to lower density regions (higher flux).  The specific values are the negative of the delta flux ($-\delta$). }\label{fig:slice100}
\end{figure*}

By fixing a $Z$, which is analogous to fixing a redshift, the predictions can be compared visually by reviewing the resulting 2D maps.  Though heuristic, this comparison is useful because it provides a clear and accessible picture of which features are revealed (missed) with more (fewer) LOS.  Figure~\ref{fig:slice1000} displays a slice 
of the predicted values for $S_{hr}$, $S_{1000}$, and $S_{1000hr}$; Figures~\ref{fig:slice200} and \ref{fig:slice100} display slices for $S_{200}$ and $S_{200hr}$, and $S_{100}$ and $S_{100hr}$, respectively.  The slices are each at the same fixed $Z$ and are directly comparable, which highlights the degree to which the datasets with fewer LOS can pick up the features present in the high-resolution dataset.  The high-resolution prediction is displayed in Figure~\ref{fig:slice1000}a, and it is interesting to compare it to $S_{1000}$ in Figure~\ref{fig:slice1000}b, $S_{200}$ in Figure~\ref{fig:slice200}a, and $S_{100}$ in Figure~\ref{fig:slice100}a.  As expected, $S_{1000}$ is able to pick up the smaller features (the ``clumps'' of higher density regions of HI) better than $S_{200}$ or $S_{100}$.  A similar assessment can be made between $S_{hr}$ and $S_{1000hr}$ in  Figure~\ref{fig:slice1000}c, $S_{200hr}$ in Figure~\ref{fig:slice200}c, and $S_{100hr}$ in Figure~\ref{fig:slice100}b:  the contrast between the high and low density regions for $S_{1000hr}$ appears almost identical to $S_{hr}$ while the resolution of the features diminishes with decreasing LOS, again as expected.  When comparing $S_{1000}$, $S_{200}$, and $S_{100}$ with $S_{1000hr}$, $S_{200hr}$, and $S_{100hr}$, respectively, there are a couple points to note.  First, the high-resolution counterparts, with significantly more observations, have much smaller standard errors (SE) than the LOS analog.  This is evident by comparing the predicted 2D map (center column of figures) with the plots to its left and right (prediction - SE and prediction + SE, respectively).  Also, $S_{1000hr}$, $S_{200hr}$, and $S_{100hr}$ have a smoother appearance with clearer features than the LOS predictions.  This is likely the result of both having more observations available for the predictions along with those observations being evenly dispersed within the simulation cube.
\\

\begin{figure*}
  \centering
 \includegraphics[width=2.25in]{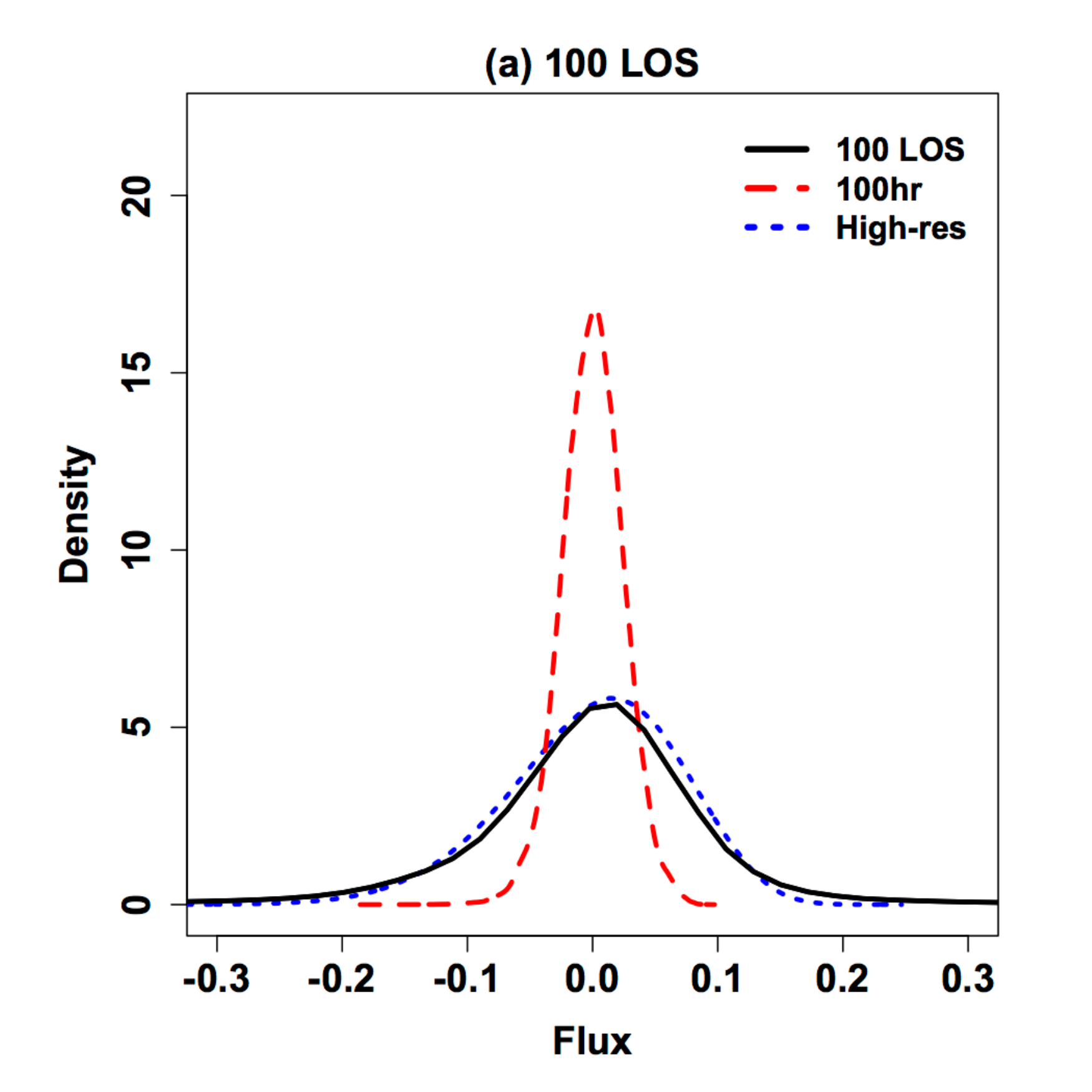}   \includegraphics[width=2.25in]{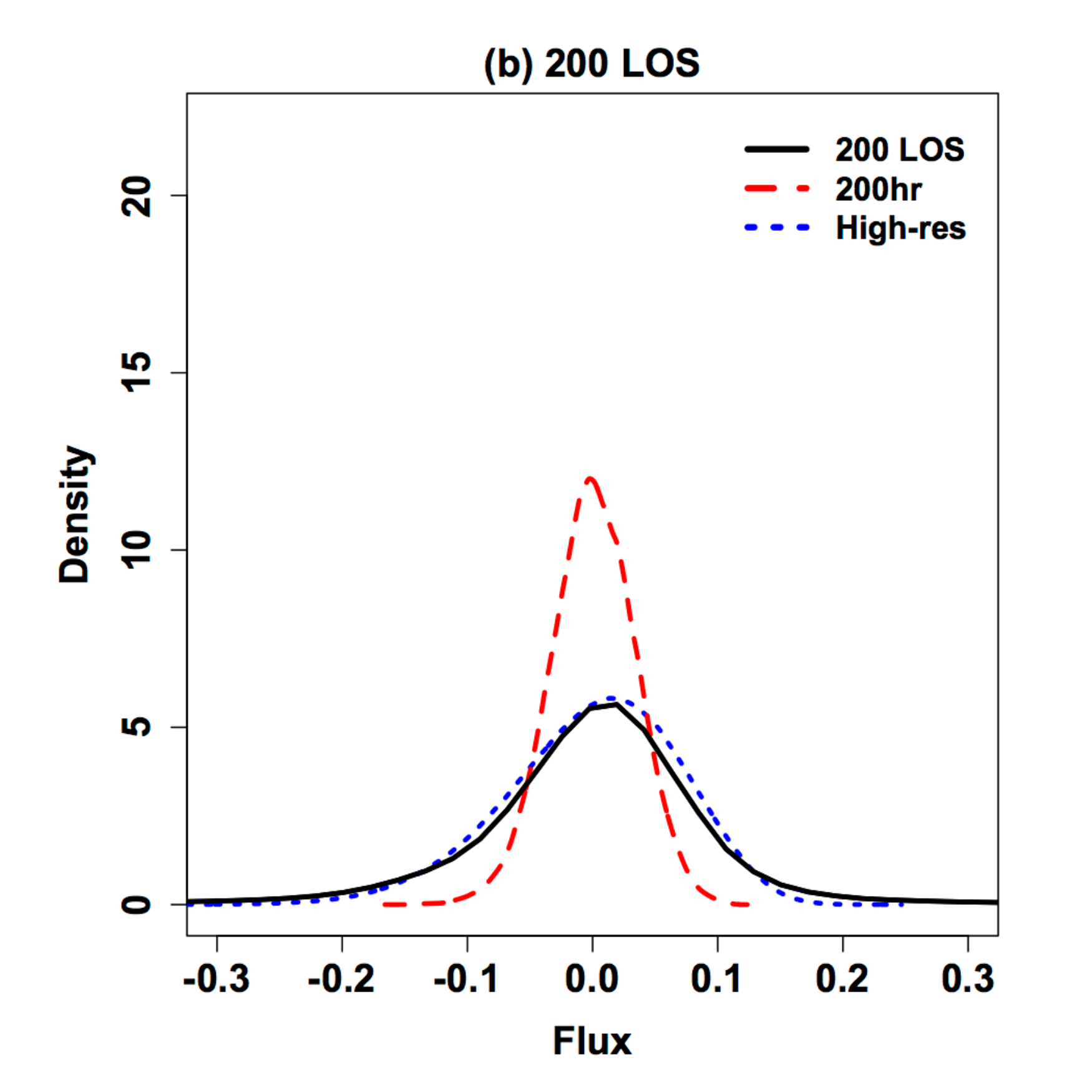}   \includegraphics[width=2.25in]{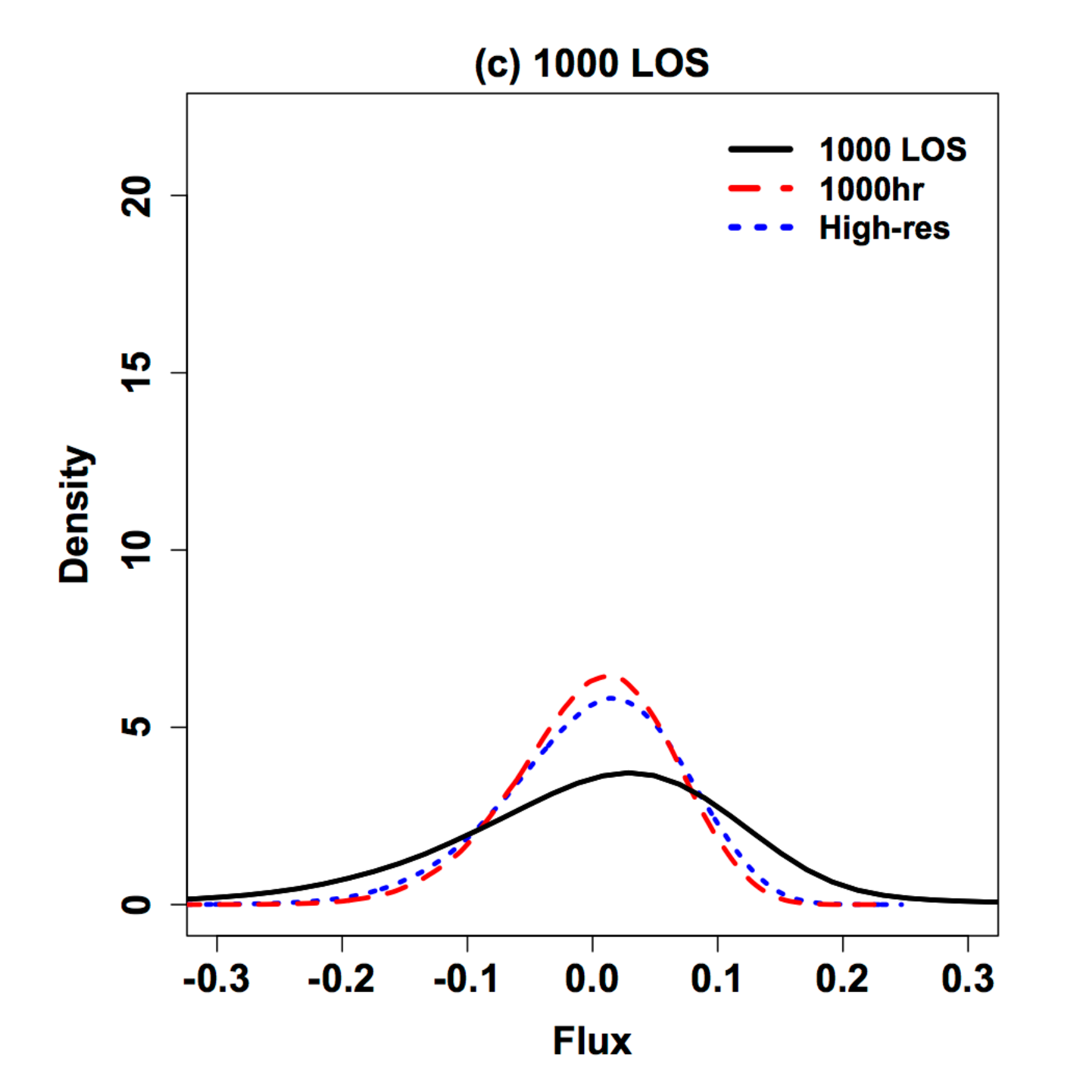} 
  \caption{The solid black lines are the PDFs for the (a) 100 LOS dataset, (b) 200 LOS dataset, and (c) 1000 LOS datasets.  The blue dotted lines the PDF for the high-resolution dataset.  The red dashed lines are the PDFs for $100hr$, $200hr$, and $1000hr$.  The PDFs are the kernel density estimates of the predicted values for the specified datasets.}
\label{fig:pdf_los}
\end{figure*}

\noindent {\bf Density estimation.} 
Comparing a summary of the distribution of predicted values via estimated PDFs is another way to analyze the modeling methodology along with assessing the changes in prediction across a varying number of LOS.  The PDFs are estimated using a Nadaraya-Watson kernel density estimator, which creates a certain smoothed histogram of the predicted maps.  This is accomplished by vectorizing the predicted values (the delta-fluxes) across $X$, $Y$, and $Z$, then applying a kernel density estimator to this vector.  Note that this eliminates any spatial information in the data.  Figure~\ref{fig:pdf_los} displays the estimated PDFs for the varying number of LOS.  The PDFs for $S_{100}$ and $S_{200}$ are closer to the PDF of $S_{hr}$ than $S_{100hr}$ and $S_{200hr}$, respectively.  This is because the latter have many more observations that are used in the local fit, which reduces the variability of the predicted delta-fluxes resulting in narrower PDFs.  The 1000 LOS dataset has a smoothing parameter closer to the smoothing parameter used for $S_{hr}$ so, when applied to the full high-resolution dataset,  it makes the predicted values for $S_{1000hr}$ very similar to the predicted delta-fluxes of $S_{hr}$.  This is also evident from the color bars for the slices of Figure~\ref{fig:slice1000}:  the range of values of Figure~\ref{fig:slice1000}c ($S_{1000hr}$) is closer to the range of values of Figure~\ref{fig:slice1000}a ($S_{hr}$) than Figure~\ref{fig:slice1000}b ($S_{1000}$).
\\

\noindent {\bf Local minima and maxima.}
One way to get a rough sense of the ``patchiness'' of the predicted maps is to count the number of local minima and maxima.  A predicted delta-flux is considered a local minimum (maximum) if it is smaller (larger) than all the other predicted values immediately surrounding it.  Periodic boundary conditions were used in the simulation design and, hence, also used in finding these local extremes.  Table~\ref{tab:modes} displays the count of local minima and maxima for the predicted maps.  The high-resolution counterparts of the 100, 200, and 1000 LOS predictions have fewer modes because more smoothing is applied due to the larger number of observations used in the local neighborhoods.  The number of local modes also provides a sense of the resolution of the features that are picked-up as the number of LOS varies:  it is expected that increasing the number of LOS increases the number of local modes.  However, it is interesting to note that the 1000 LOS dataset has more local modes than the full, high-resolution dataset.  This may be due to a balance of two facts:  (i) fewer observations are used in the local neighborhood for $S_{1000}$ compared to $S_{1000hr}$ or $S_{hr}$ allowing for more extreme values due to less smoothing, and (ii) a small enough bandwidth so that local features are appearing leading to several minima and maxima.
\\

\begin{center}
\begin{table}
\begin{tabular}{ccc}
{\bf Sample} & {\bf Local Max.} & {\bf Local Min.} \\
\hline
$S_{100}$    & 901    & 832 \\
$S_{100hr}$    & 350    & 365 \\
$S_{200}$   & 1,867   & 1,750 \\
$S_{200hr}$    & 439    & 424 \\
$S_{1000}$   & 9,823   & 8,435 \\
$S_{1000hr}$   & 2,401   & 2,166 \\
$S_{hr}$   & 4,398   & 3,809 \\
Raw data & 164,172 & 159,098 \\
\hline
\end{tabular} \caption{Count of local maxima and minima providing a rough measure of the ``patchiness'' of the predicted maps.} \label{tab:modes}
\end{table}
\end{center}

\noindent {\bf Correlation function.}

\begin{figure}
   \centering
\includegraphics[width=3.25in]{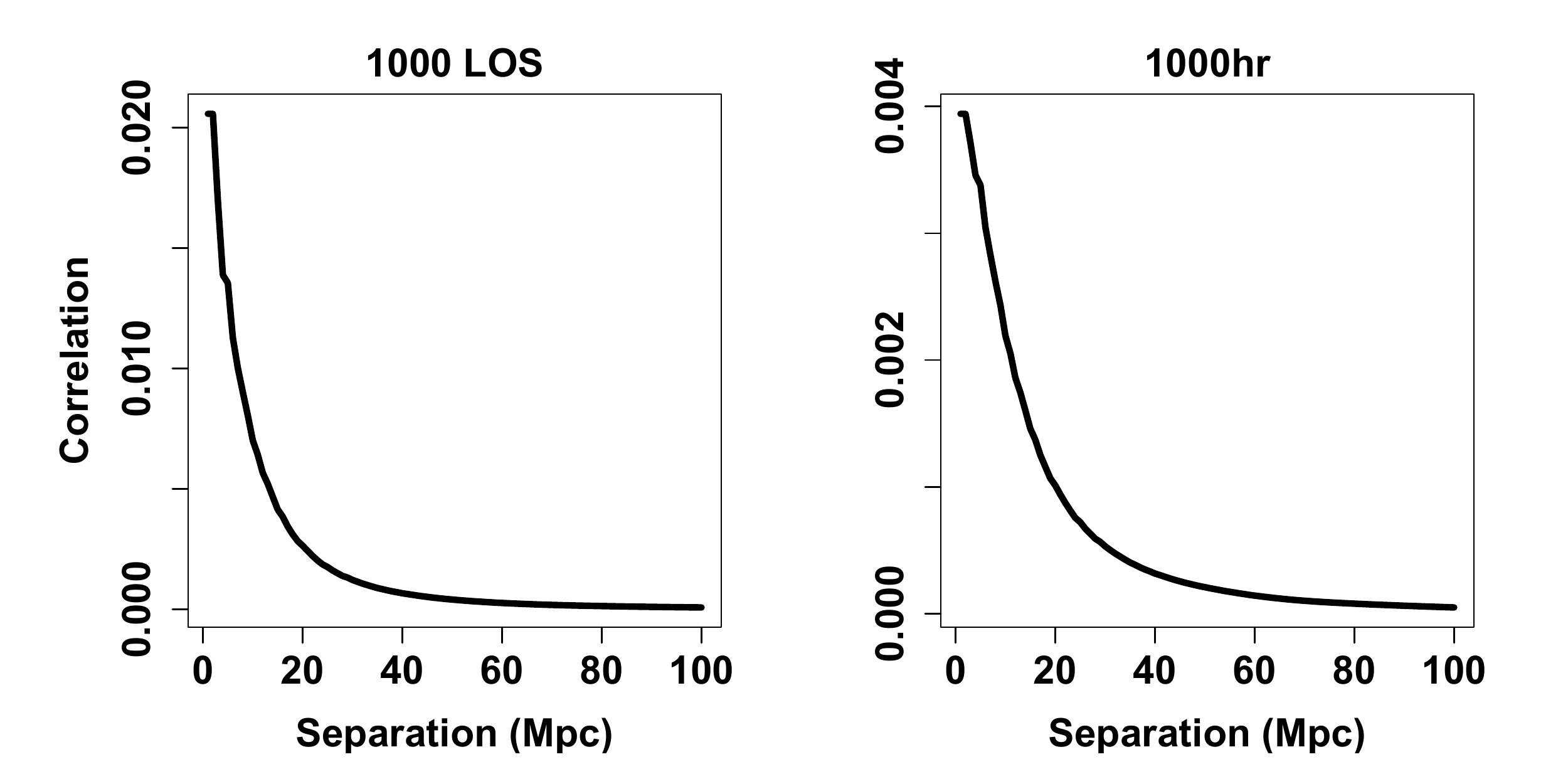} 
\includegraphics[width=3.25in]{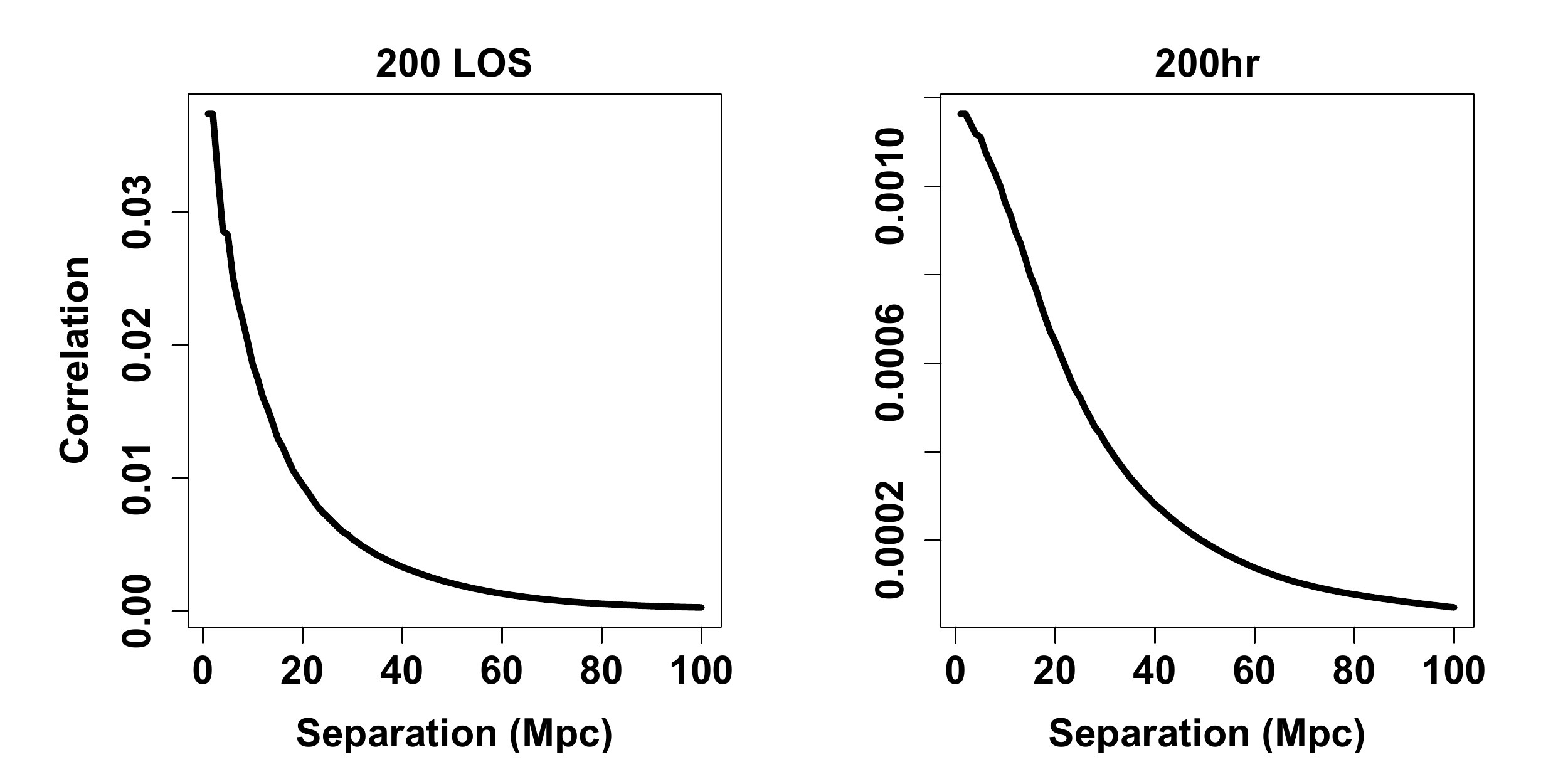} 
\includegraphics[width=3.25in]{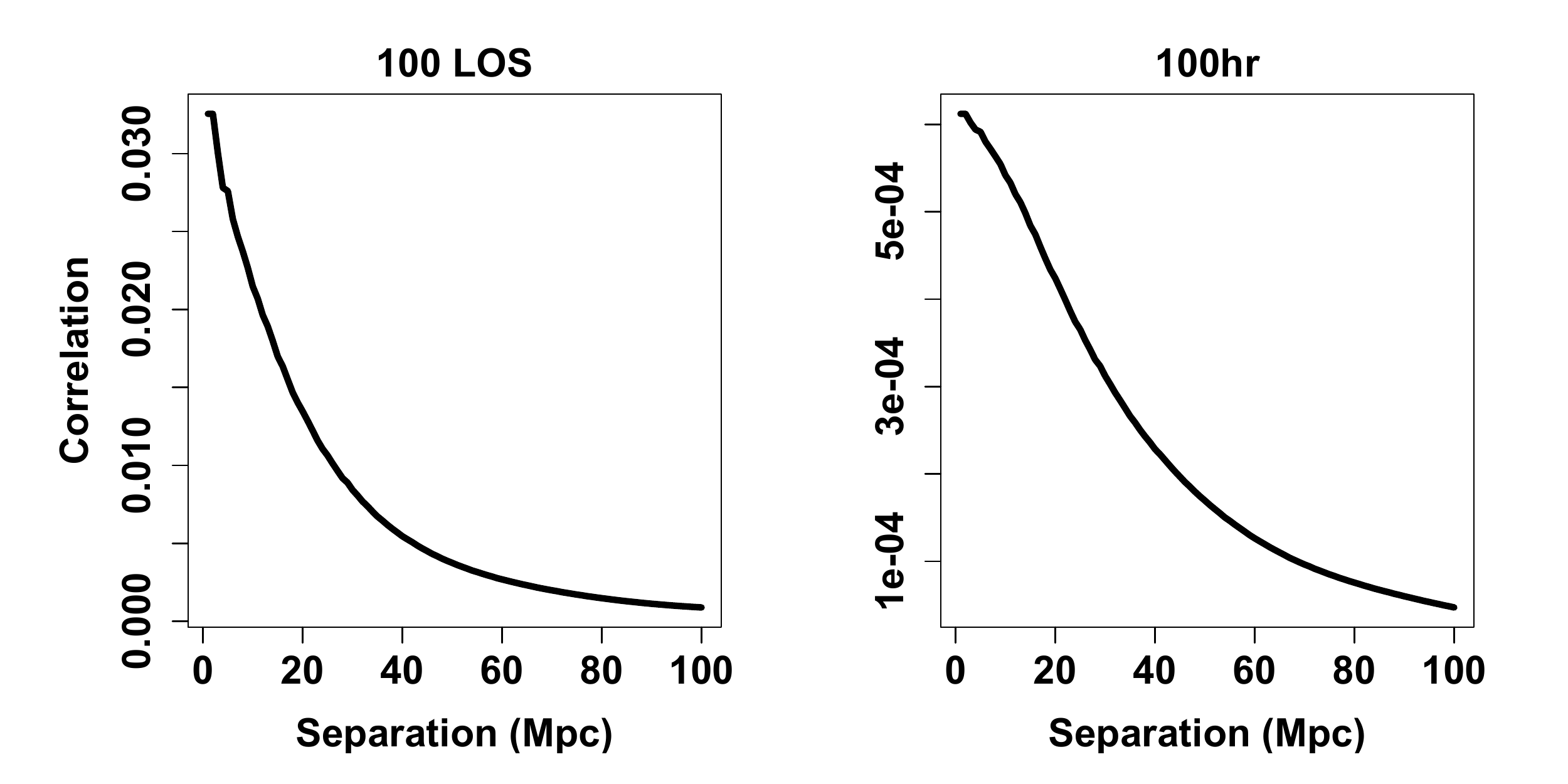} 
   \caption{Correlation functions for LOS datasets:  $S_{1000}$ and $S_{1000hr}$ (top row); $S_{200}$ and $S_{200hr}$ (middle row); $S_{100}$ and $S_{100hr}$ (bottom row).}
 \label{fig:corr}
 \end{figure}

\begin{figure}
   \centering
\includegraphics[width=3.5in]{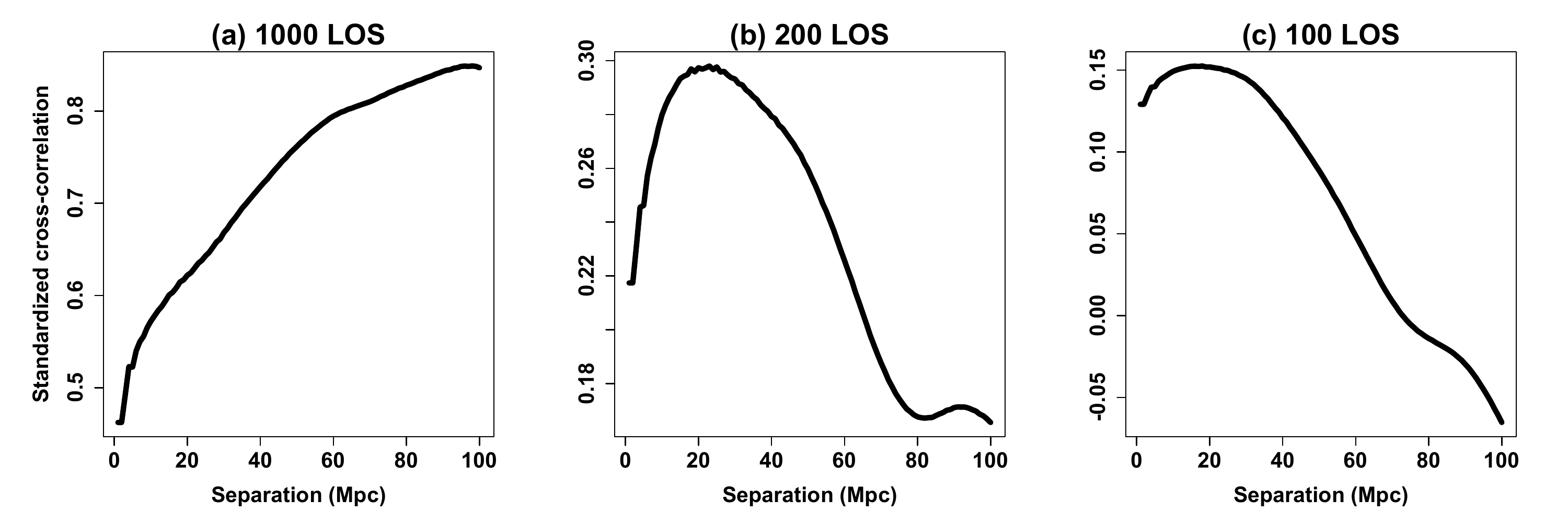} 
   \caption{The standardized cross-correlation between (a) $S_{1000}$ and $S_{1000hr}$,  (b) $S_{200}$ and $S_{200hr}$, and (c) $S_{100}$ and $S_{100hr}$.}
 \label{fig:cross}
 \end{figure}

The correlation function is often used to capture the spatial features of a region.  The correlation functions for $S_{1000}$, $S_{1000hr}$, $S_{200}$, $S_{200hr}$, $S_{100}$, and $S_{100hr}$ are displayed in Figure~\ref{fig:corr}.  Comparing the correlation functions of $S_{200}$ with $S_{200hr}$, and $S_{100}$ with $S_{100hr}$, it is evident that the the correlation drops off faster with $S_{200}$ and $S_{100}$ than their HR counterparts.  The discrepancies between these correlation functions can be quantified using a standardized cross-correlation.  Let $A_1$ and $A_2$ be the correlation functions for two different datasets, and $C_{12}$ be their cross-correlation function, then the standardized cross-correlation is $C_{12}/\sqrt{A_1\cdot A_2}$.  Figure~\ref{fig:cross} displays the standardized cross-correlation functions for $S_{1000}$ and $S_{1000hr}$,  $S_{200}$ and $S_{200hr}$, and $S_{100}$ and $S_{100hr}$.  While the standardized cross-correlation between $S_{1000}$ and $S_{1000hr}$ increases to 1 as the separation increases (as desired), the other two standardized cross-correlation functions imply a significant disparity in their spatial behavior.  This is likely due to the difference in the number of observations that are used in the local fit:  there are 118,139 \emph{more} observations in the local neighborhoods for $S_{100hr}$ over $S_{100}$; about 45,282 more observations for $S_{200hr}$ over $S_{200}$; and only about 4,379 more observations in the local neighborhoods for $S_{1000hr}$ over $S_{1000}$.  These differences are expected to materialize in the correlation function as is evidenced in the noted figures. 
\\

\noindent {\bf Persistent homology.}
Persistent homology is a tool in topological data analysis that could prove to be useful in cosmology as suggested in \cite{Sousbie2011}, \cite{SousbieEtAl2011}, and \cite{WeygaertEtAl2013}.  An introduction to persistent homology can be found in \cite{EdelsbrunnerHarer2008} and \cite{Carlsson2009}, and some advances in statistical inference for persistent homology can be found in \cite{BalakrishnanEtAl2013}. 

Trying to resolve the desire to understand the connectedness and the shape of the object of interest with the fact that data are collect discretely produces significant challenges in topological data analysis.   One way in which persistent homology overcomes this issue by considering upper level sets of the estimated function, $L_{\lambda} = \{x \mid \hat{f}(x) > \lambda\}$, which are the inputs to the function such that the function value at that input is larger than $\lambda$, where $\lambda$ can be considered a tuning parameter.  As $\lambda$ decreases, certain topological features will appear (e.g. loops) and then disappear.  The birth and death of these features provide a topological signature of the object, which can be displayed in a plot with the $\lambda$ value of the birth and death of each feature plotted as a point;  this plot is called a \emph{persistence diagram}.  Intuitively, points further from the 45 degree line are topological features that live longer as the tuning parameter decreases, and may be considered real features while those close to the line may be considered topological noise (though this is not always the case).

\begin{figure} 
  \centering
     \includegraphics[width=3.5in]{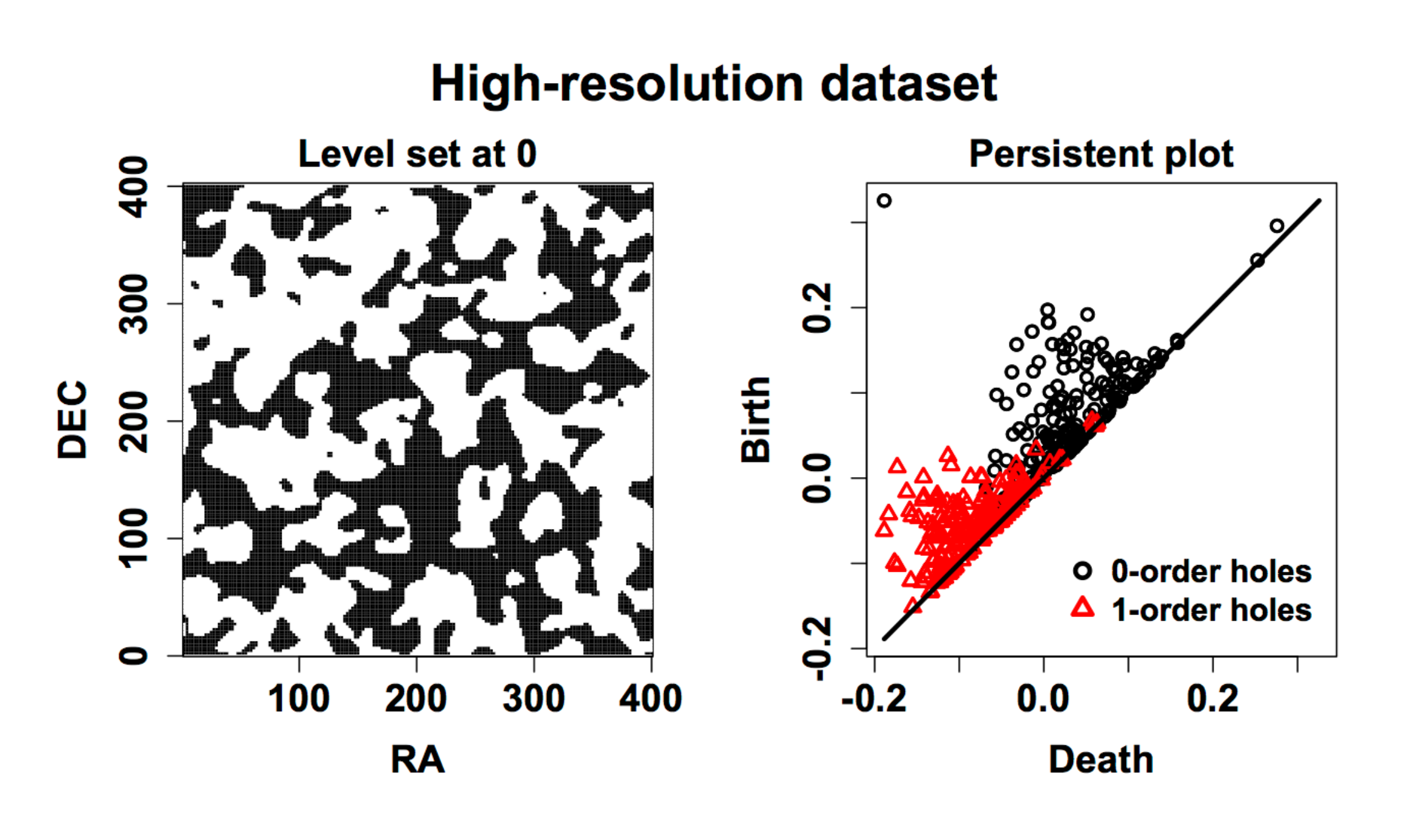} 
  \caption{The persistence diagram (left) for a slice of $S_{hr}$, and the upper level set at 0 (right) corresponding to the top centre image in Figure~\ref{fig:slice1000}.  The black circles represent 0-order holes (connected components) and the red triangles are 1-order holes (loops).}
\label{fig:phom0}
\end{figure}

For the Lyman-alpha forest data, the upper level sets are the locations of the observations, $(X, Y,  Z)$ such that the corresponding smoothed estimate of the delta-flux, $\delta$, are greater than $\lambda$.  The left plot in Figure~\ref{fig:phom0} displays the upper level set at $\lambda = 0$ for a slice of $S_{hr}$ corresponding to the predicted slice in Figure~\ref{fig:slice1000}a, and the right plot is the persistence diagram for that slice.  The black circles and red triangles represent the birth and death of 0- and 1- order holes, respectively, where 0-order holes are connected components and 1-order holes are loops.  The persistence diagram shows that as $\lambda$ decreases, the points in the upper level set are becoming connected, eventually forming many loops and ultimately becoming one connected component.

\begin{figure*} 
\centering
\includegraphics[width=2.15in]{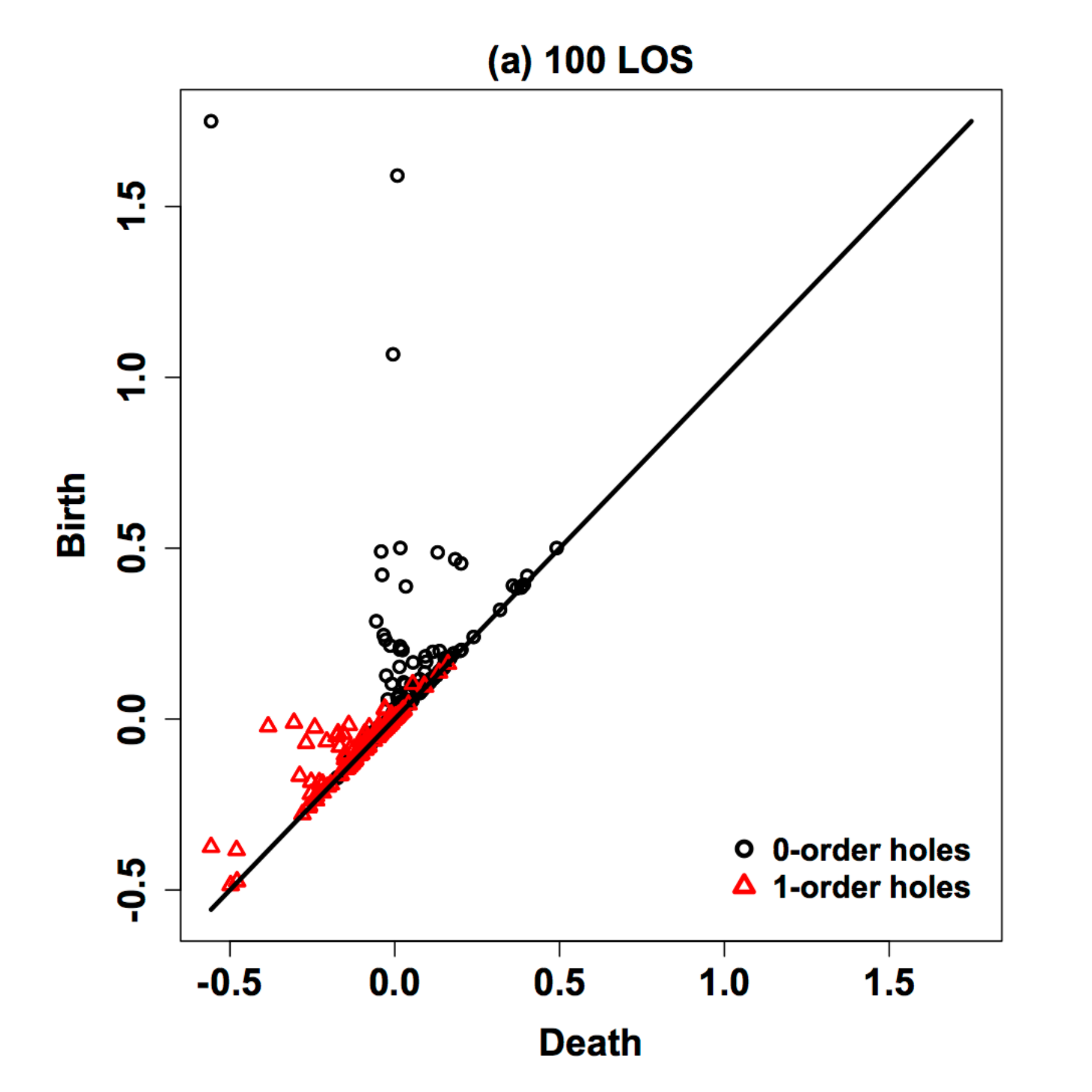} 
\includegraphics[width=2.15in]{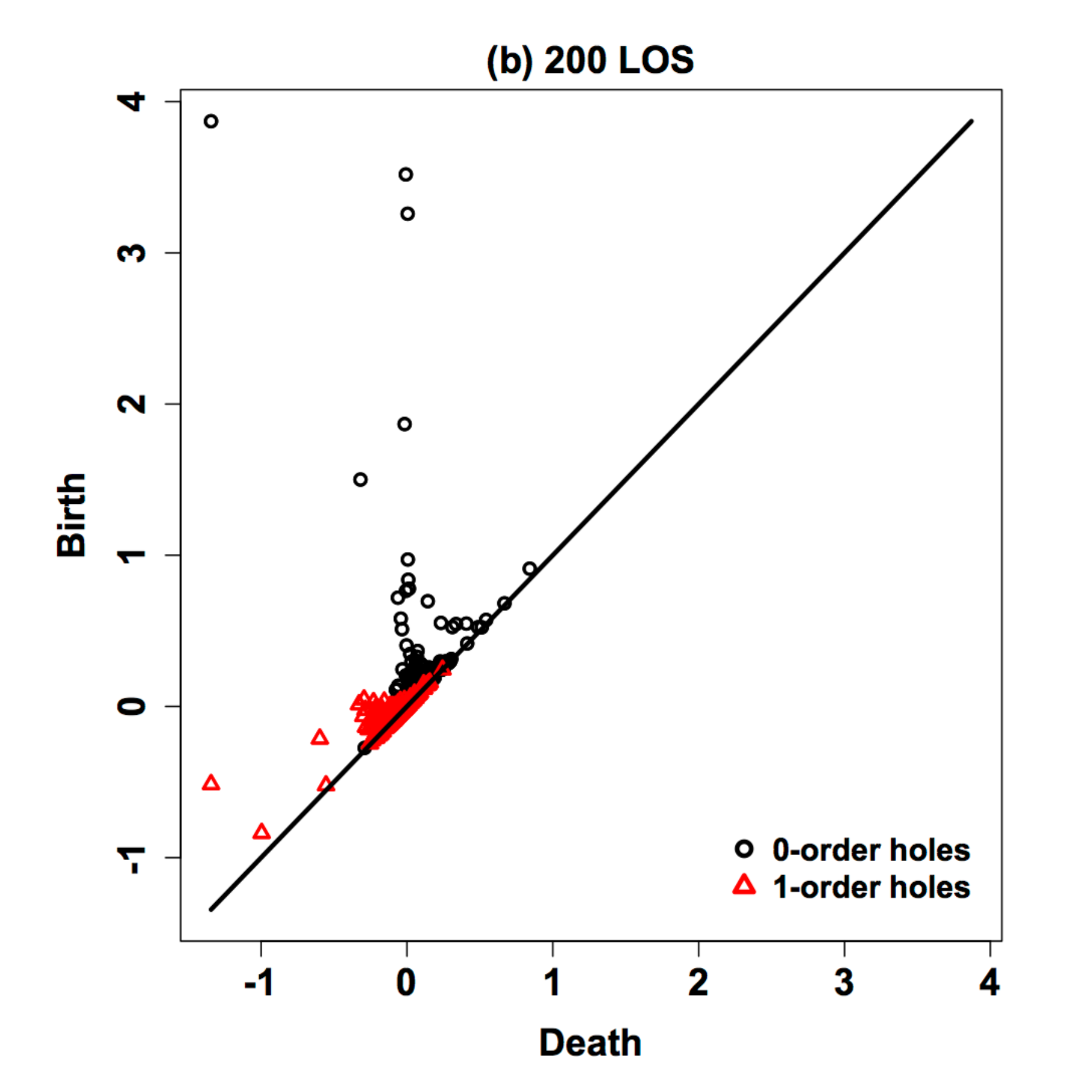} 
\includegraphics[width=2.15in]{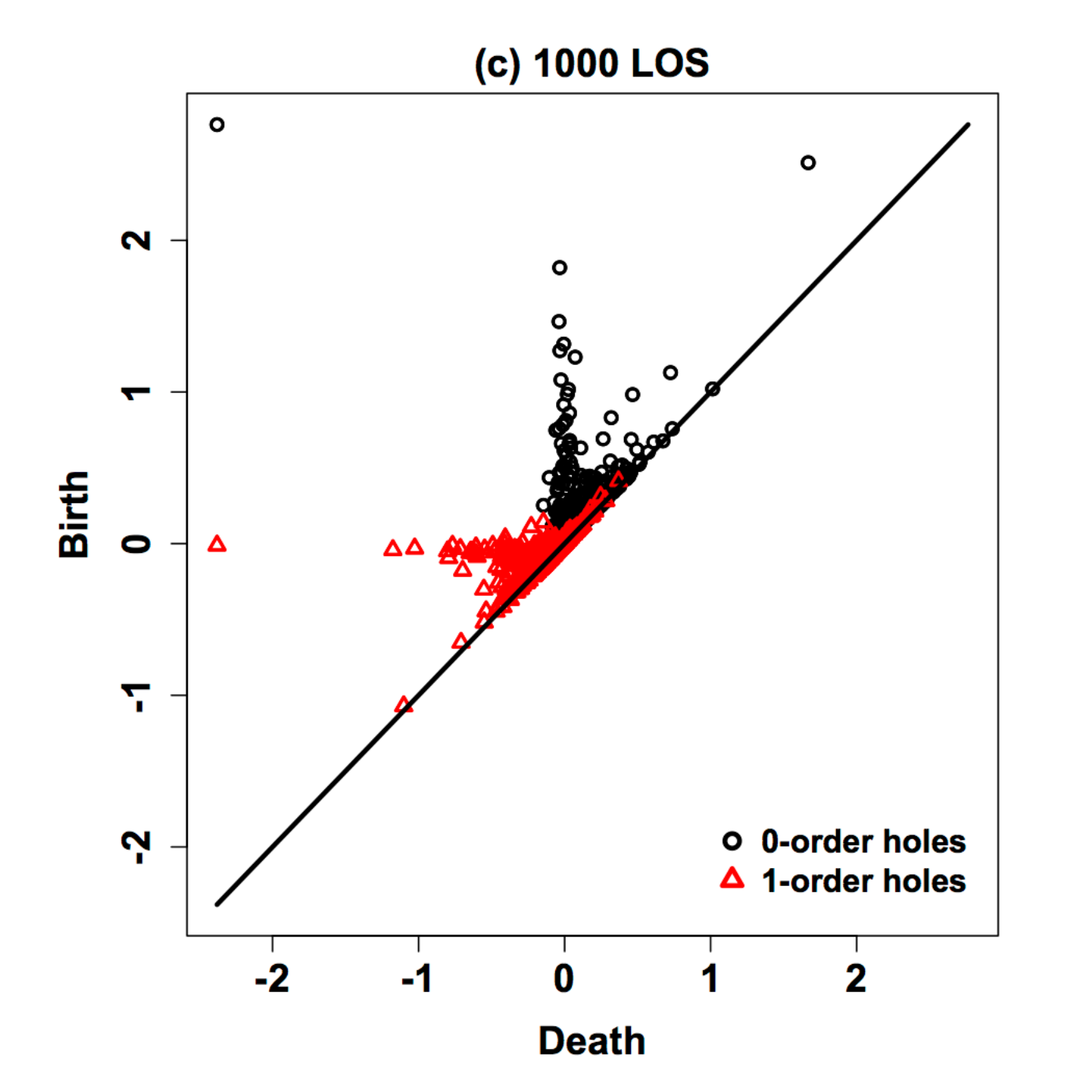} 
\includegraphics[width=2.15in]{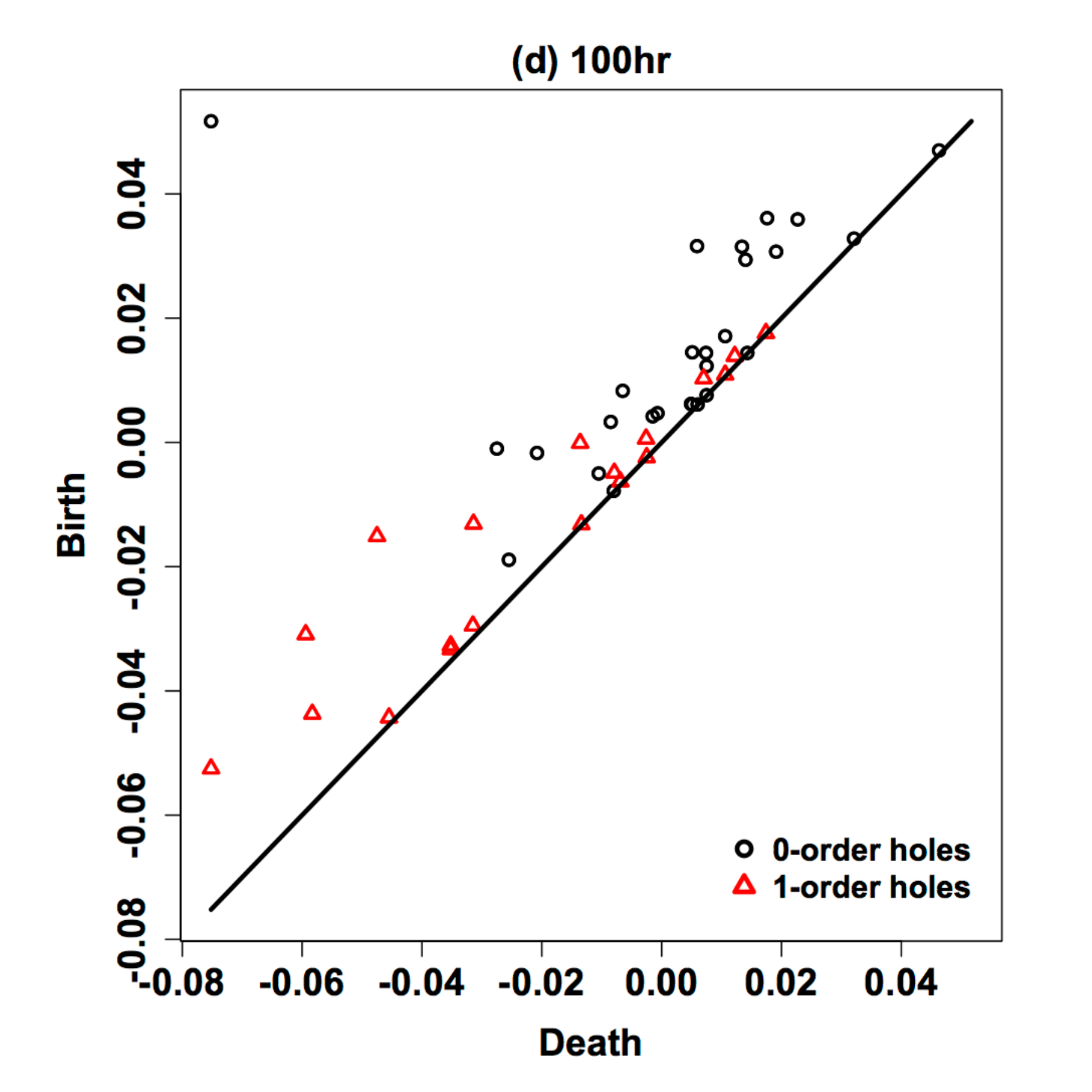} 
\includegraphics[width=2.15in]{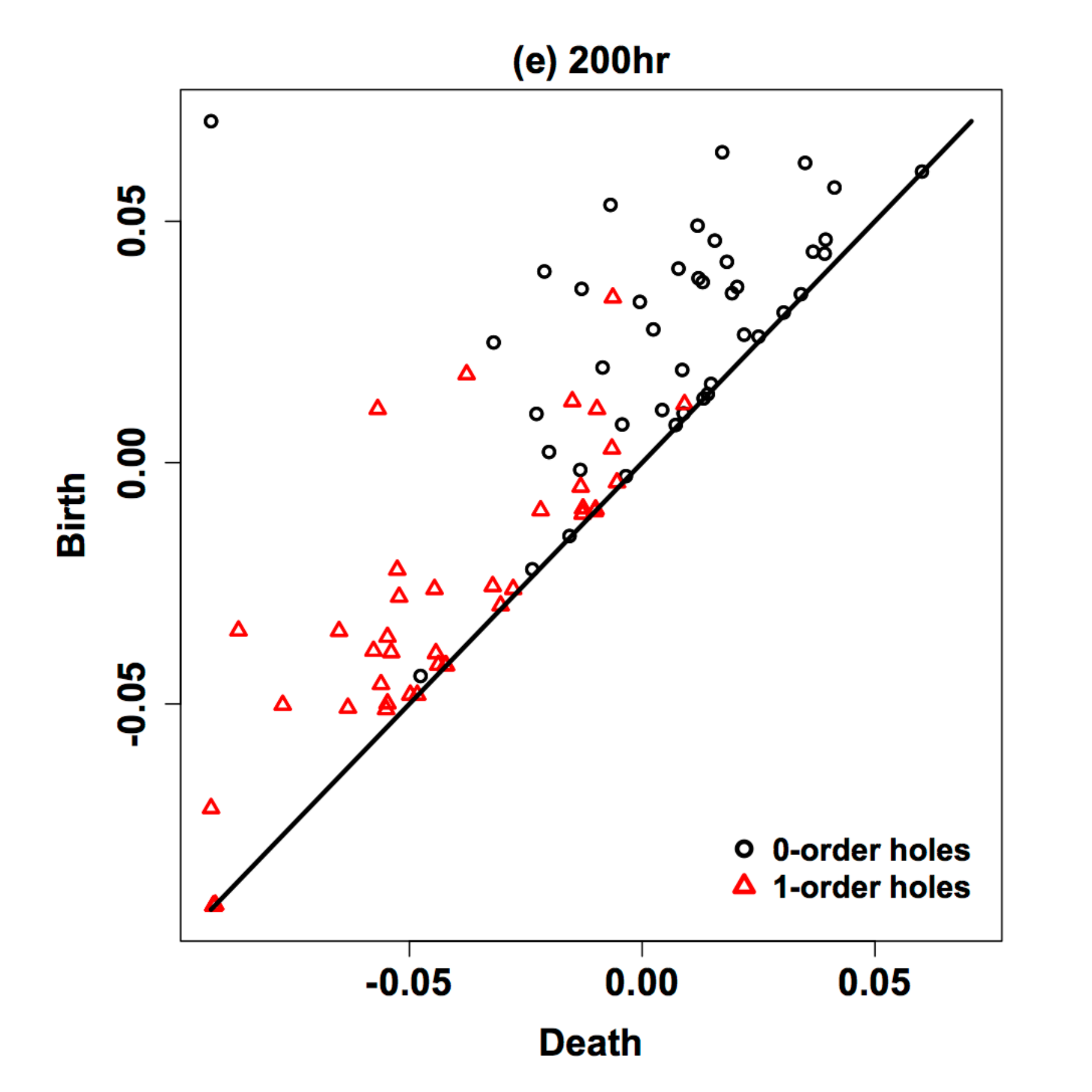} 
\includegraphics[width=2.15in]{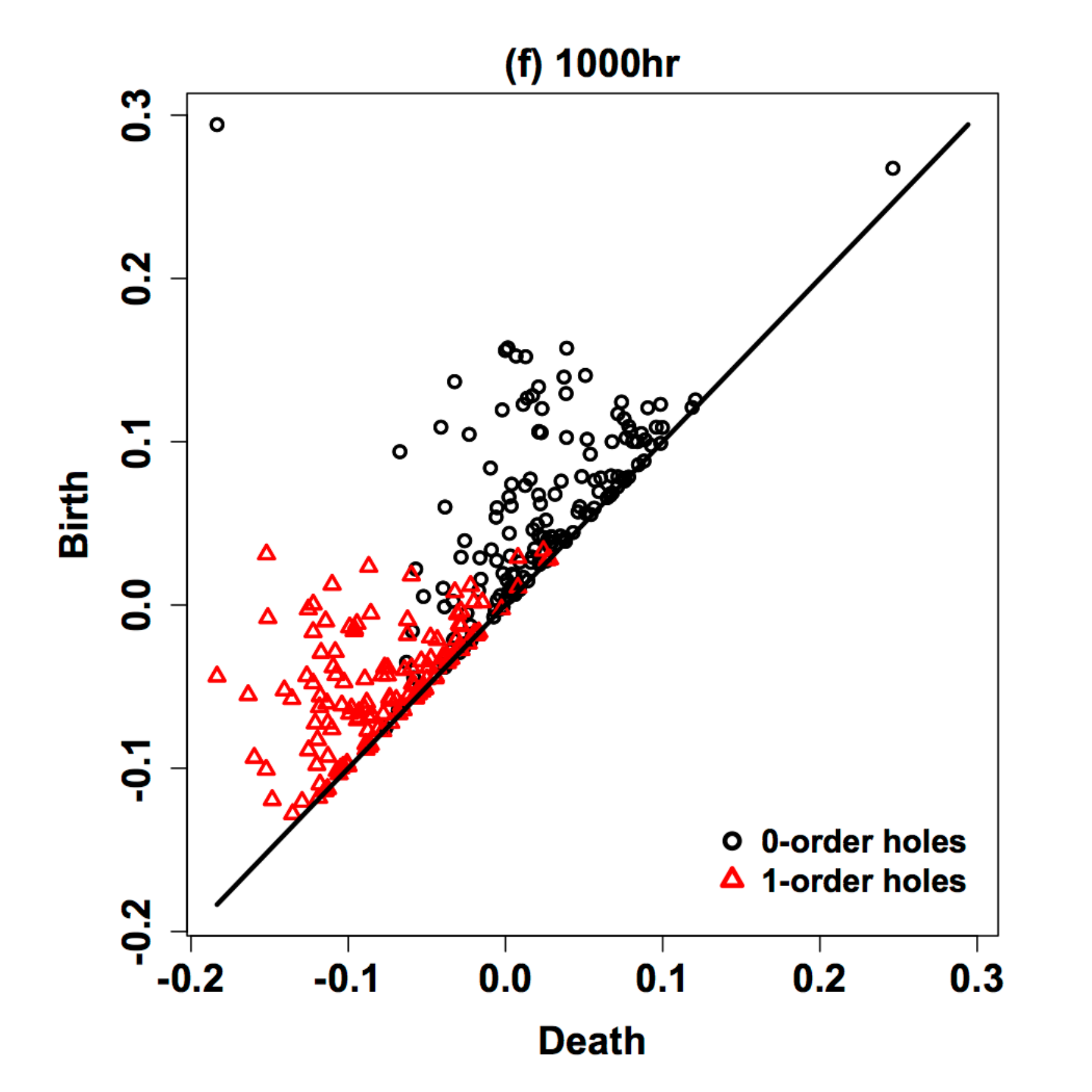} 
\caption{Persistence diagrams for a slice of (a) $S_{100}$, (b) $S_{200}$, (c) $S_{1000}$, (d) $S_{100hr}$, (e) $S_{200hr}$, and (f) $S_{1000hr}$.  The black circles represent 0-order holes (connected components) and the red triangles are 1-order holes (loops).}
\label{fig:phom1}
\end{figure*}

Figure~\ref{fig:phom1} displays the persistence diagrams for the $S_{100}$, $S_{200}$, and $S_{1000}$ along with persistence diagrams for $S_{100hr}$, $S_{200hr}$, and $S_{1000hr}$.  A feature shared among all these datasets is that, as $\lambda$ decreases, the number of connected components decreases as expected, but interestingly, the connections are forming loops (1-order holes).  Figures~\ref{fig:phom1}d and \ref{fig:phom1}e suggest by the proximity of the points to the 45 degree line that the topological features were lost due to increased smoothing.   Figure~\ref{fig:phom1}c has an interesting feature: there is a vertical line of 0-order holes and a horizontal line of 1-order holes (the vertical line is also somewhat evident in Figures~\ref{fig:phom1}a and \ref{fig:phom1}b).  This is likely a result of the smoothing methodology and the non-regular design due to random sampling of QSO LOS since these features are not appearing in the persistence diagrams for $S_{100hr}$, $S_{200hr}$, and $S_{1000hr}$, which have evenly distributed observations.

Though the current comparison of these persistence diagrams is heuristic, more formal comparisons are forthcoming.

\section{Application to a sample of BOSS data}
\begin{figure}
  \centering
     \includegraphics[width=2.5in]{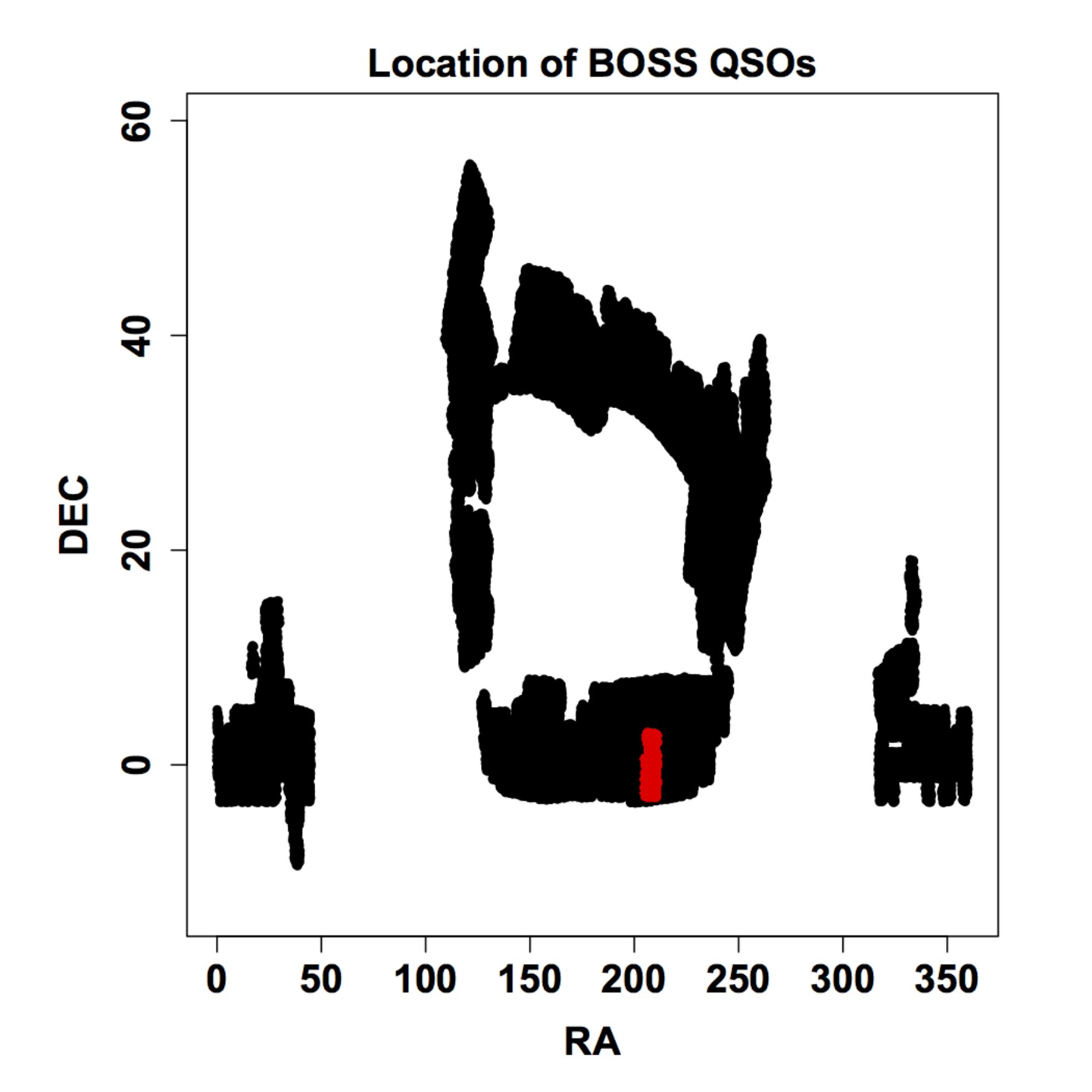} 
  \caption{RA and DEC of BOSS DR9 QSOs.  The red region (RA:  205 to 211, DEC:  -3 to 3) is the region selected for the analysis.}
\label{fig:boss_map}
\end{figure}

\begin{figure} 
  \centering
     \includegraphics[height=1.5in]{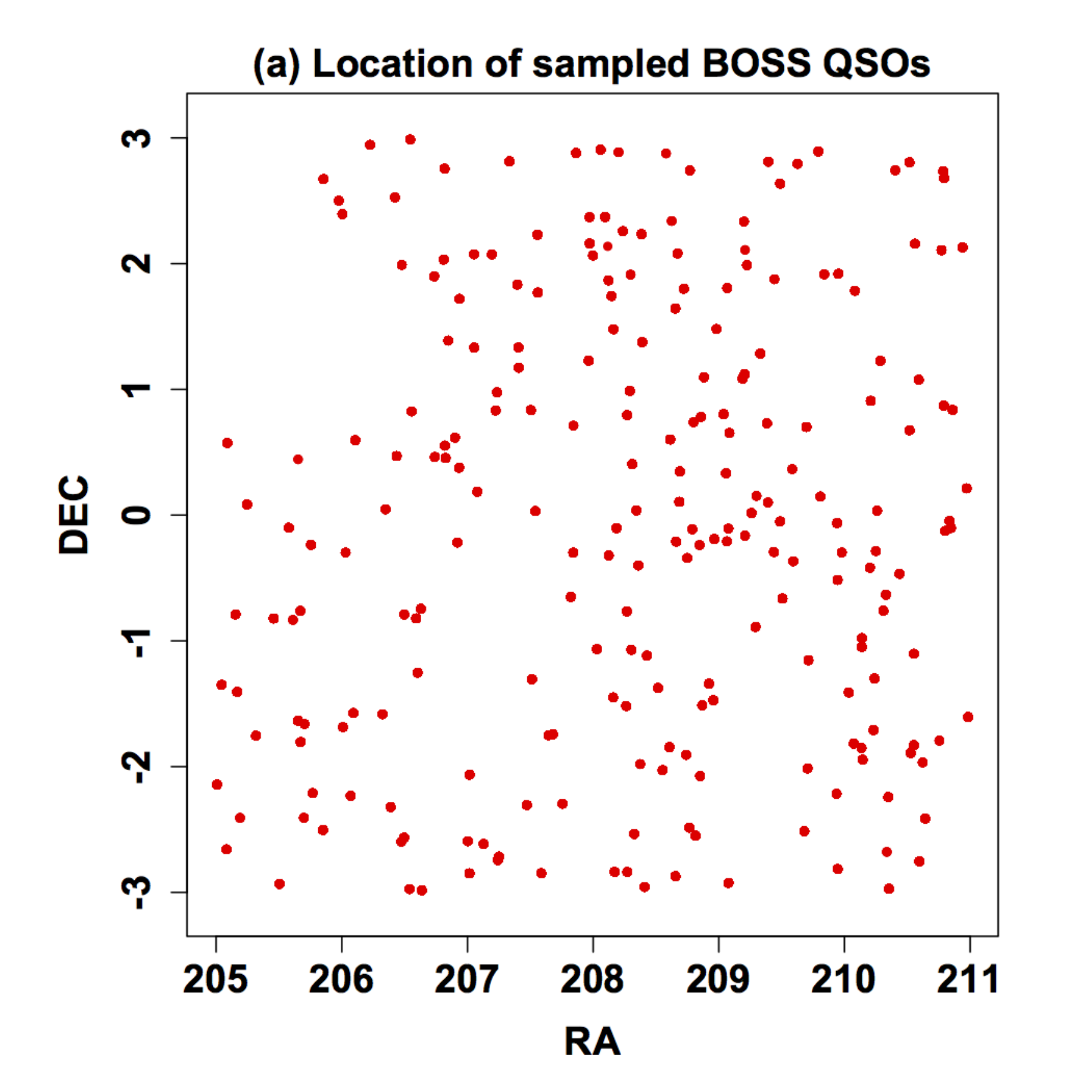} \qquad
     \includegraphics[height=1.5in]{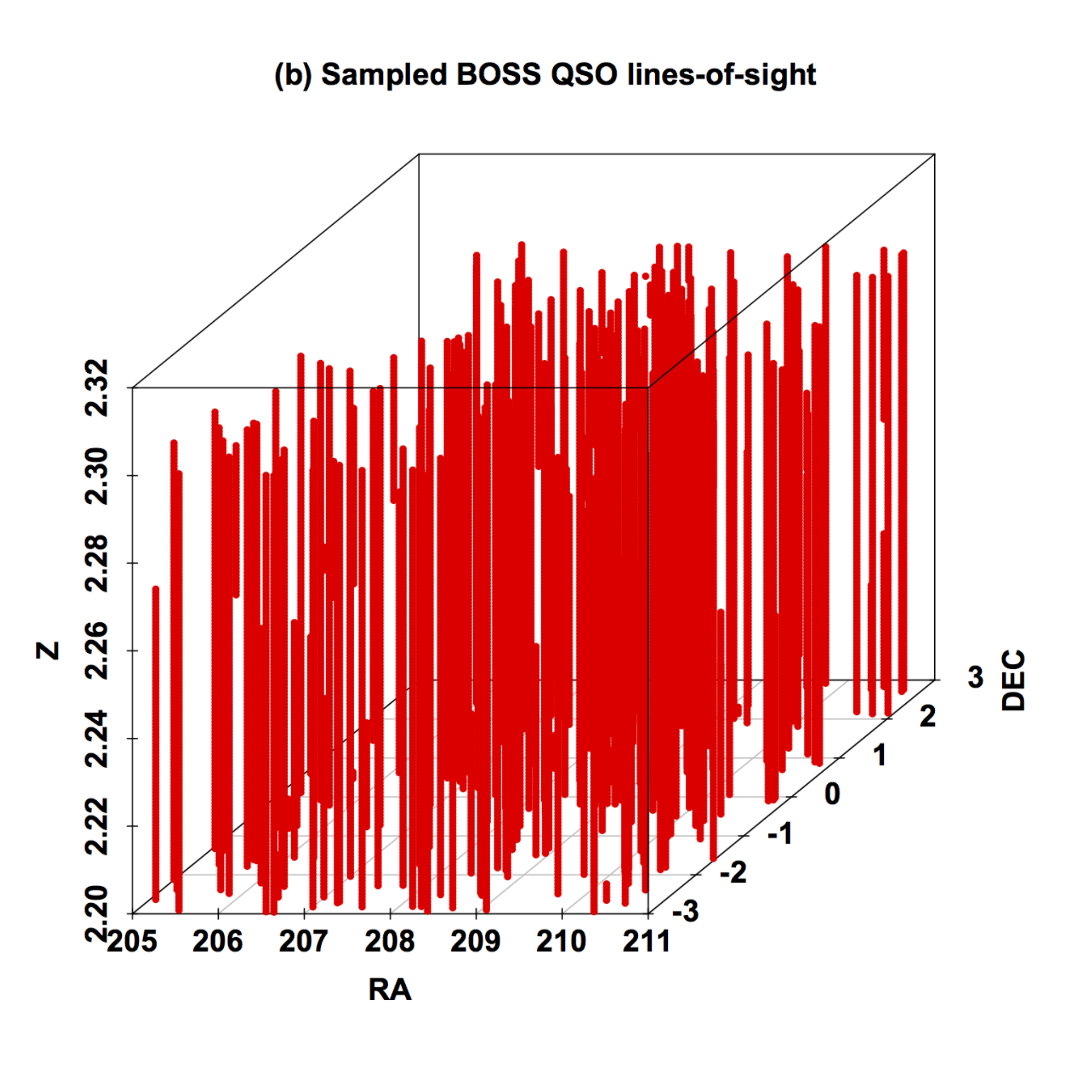} 
  \caption{(a) RA and DEC of sampled BOSS QSOs (RA:  205 to 211, DEC:  -3 to 3) and (b) 3D lines of sight of sampled QSOs (Z: 2.2 to 2.3).}
\label{fig:sampled_quasars}
\end{figure}

The previous section demonstrated the performance of local polynomial smoothing in producing a 3D map of a (400 h$^{-1}$Mpc)$^3$ simulation cube with varying numbers of LOS.  In this section we apply the same methodology to a subset of the Lyman-alpha forest data from BOSS SDSS DR9 (\citealt{LeeEtAl2013}) that is roughly the same volume as the simulation cube of the previous section.  The BOSS Lyman-alpha forest data from DR9 has 54,468 QSO spectra at redshifts greater than 2.15 with absorption redshifts between 2.0 and 5.7.  Figure~\ref{fig:boss_map} displays the RA and DEC of these QSOs.  The region selected for this analysis contains 234 QSO with 24,596 measurements between an RA of 205 and 211, a DEC between -3 and 3, with redshifts between 2.2 and 2.3.  The RA and DEC of the sampled QSOs are displayed in Figure~\ref{fig:sampled_quasars} along with a 3D view of the LOS in that region.  Much like the varying LOS data of the previous section, the LOS in the volume selected are not evenly spaced.  Unlike the simulated data, not all skewers completely traverse the cube, which makes the data more sparse at the top of the cube.

\begin{figure*}
  \centering
  \includegraphics[height=2.5in]{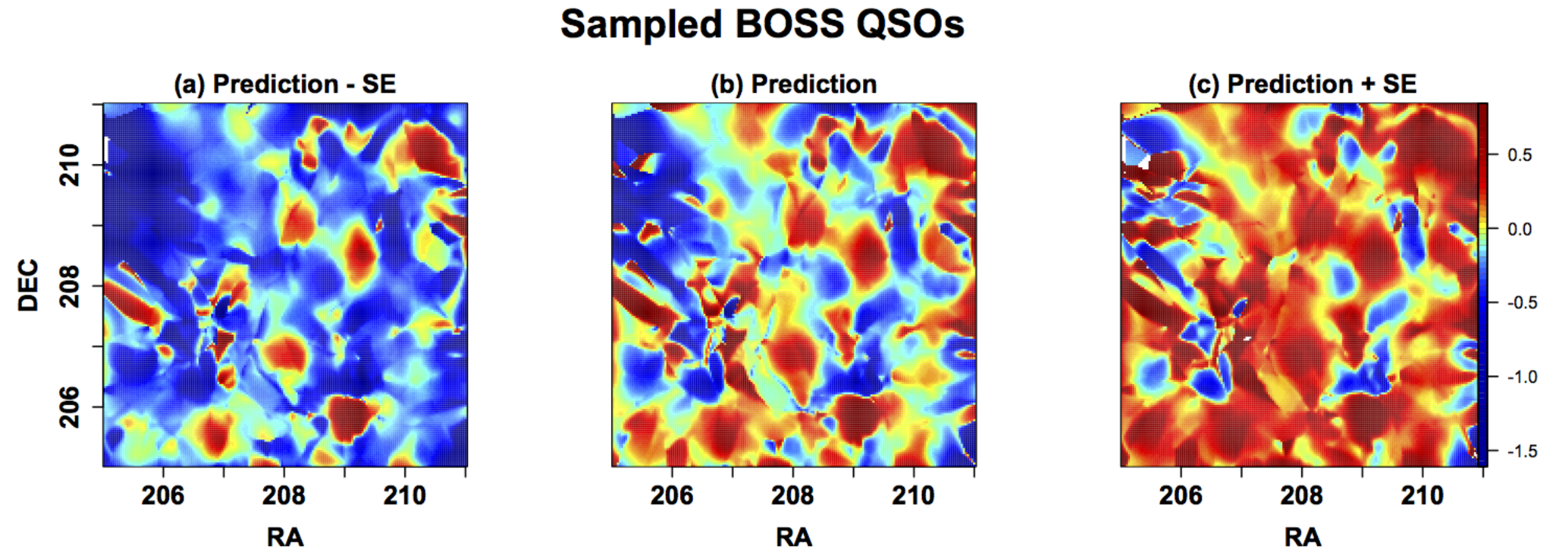} 
  \caption{Slices of the predicted values of the sample of BOSS data: (a) predicted values minus standard error (SE), (b) predicted values, (c) predicted values plus SE.  Note that the red corresponds to higher density regions (lower flux) while the blue corresponds to lower density regions (higher flux).  The specific values are the negative of the flux. }\label{fig:slice_real}
\end{figure*}

\begin{figure} 
  \centering
     \includegraphics[width=3.5in]{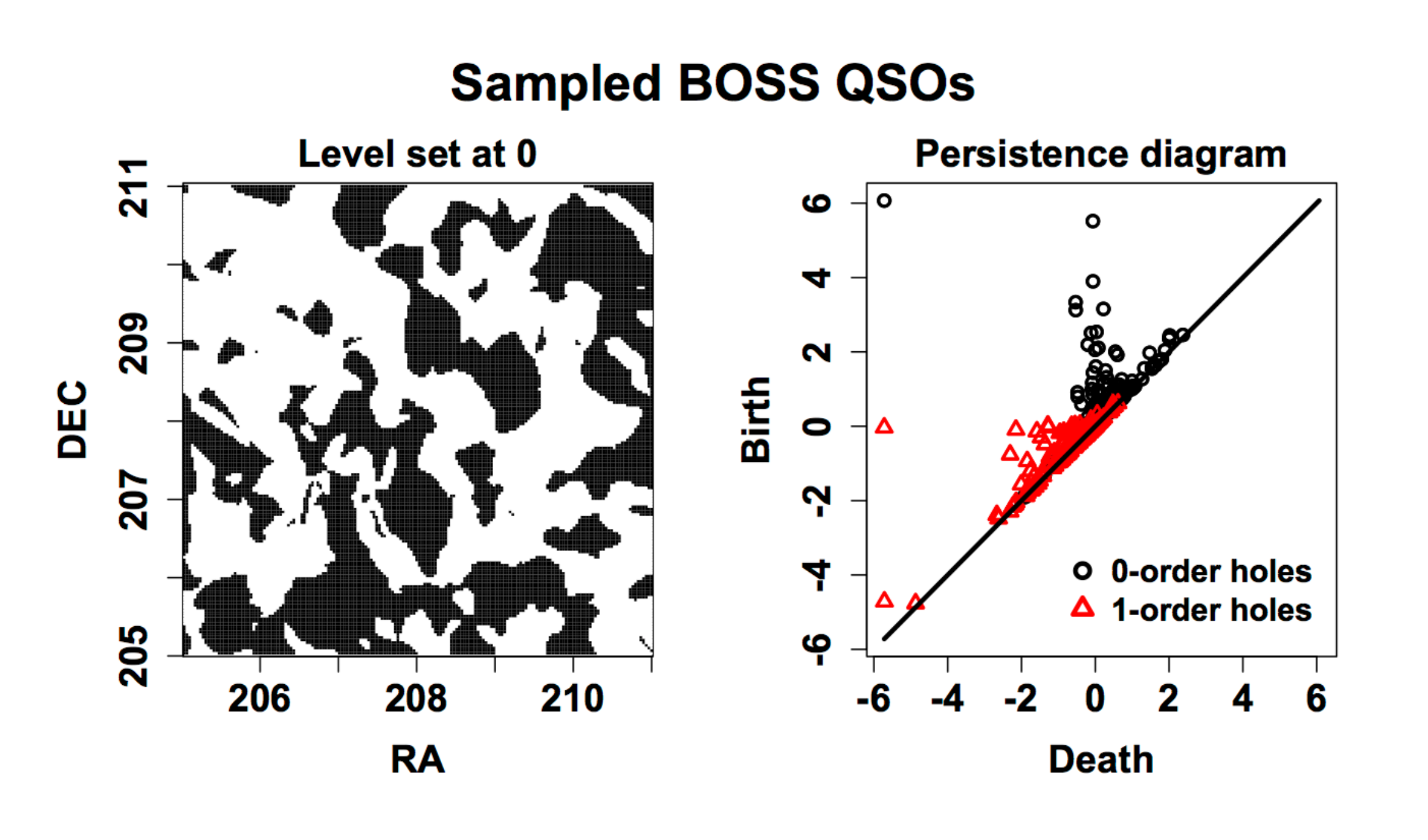} 
  \caption{The persistence diagram (left) for a slice of the sampled BOSS data and the upper level set at 0 (right) corresponding to the centre image in Figure~\ref{fig:slice_real}.  The black circles represent 0-order holes (connected components) and the red triangles are 1-order holes (loops).}
\label{fig:phom_boss}
\end{figure}

As before, the smoothing parameter was selected using GCV, and the selected value was $0.01$ resulting in 246 observations falling into each neighborhood.  A slice of the predicted region is displayed in Figure~\ref{fig:slice_real}.  Given that the selected region contains only 234 LOS, it is not surprising that the slices resemble the slices for 200 and 100 LOS displayed in Figures~\ref{fig:slice200} and \ref{fig:slice100}.  The persistence diagram of the slice is displayed in Figure~\ref{fig:phom_boss}, which also resembles the behavior of the persistence diagrams of the previous section:  as the upper level-set threshold decreases, the number of connected components die off turning into loops.

\section{Conclusions}

Modeling the IGM at high redshifts using Lyman-alpha forest data is an interesting problem --- both scientifically and statistically.  Scientifically, having an understanding of the distribution of matter at high redshifts is crucial for understanding the evolution of the Universe.  Unfortunately, data at these high redshifts are sparse and attempting to resolve the regions between observations is not straightforward.  Local polynomial smoothing is a nonparametric methodology that offers a reasonable approach for accomplishing this challenging task.  Local polynomial smoothing adapts naturally to unevenly space observations such as the location of QSOs, it eliminates certain biases present in kernel regression, and it provides standard errors for the estimates.  By comparing slices, estimated PDFs, local minima and local maxima, correlation functions, and persistence diagrams, we have captured and reviewed the changes in predictions of local polynomial smoothing as the number of LOS decreases.

There are a number of additional questions to investigate related to the proposed setting.  Noted previously, the particular design of the data is unusual in that the observations lie on almost parallel lines.  Due to this special and peculiar design, any statistical methodology employed should explicitly account for this.  For example, the local polynomial smoothing parameter $\alpha$ should have an adjustment to ensure that the neighborhood defined for estimation is not selecting an erroneous number of observations falling on the same QSO LOS potentially leading to a bias.  This issue is not unique to local polynomial smoothing, but any localized statistical methodology.  Another problem to investigate is assessing the performance when incorporating realistic error structures into the simulation data (i.e., error structures analogous to error expected in the real data).  Adding homoscedastic error to the simulation cube did not alter the results significantly. 

\section{Acknowledgments}
Jessi Cisewski was partially supported by the National Science Foundation under Grant DMS-1043903. This work was also supported by NSF awards AST1109730.
and OCI-0749212. This research
was enabled by an allocation of advanced computing resources
provided by the National Science Foundation. The large-scale computations
were performed on the Kraken facility
at the National Institute for Computational
Sciences (http://www.nics.tennessee.edu).
We thank Volker Springel and Tiziana Di Matteo for use of
the \small{P-GADGET} simulation code and the simulation data used here.
Any opinions, findings, and conclusions or recommendations expressed in this material are those of the authors and do not necessarily reflect the views of the National Science Foundation.

\bibliographystyle{mn2e}
\bibliography{$HOME/Dropbox/paper/labib.bib}

\label{lastpage}

\end{document}